\documentclass[10pt,journal,compsoc]{IEEEtran}

\usepackage[colorlinks=true,linkcolor=black,anchorcolor=black,citecolor=black,filecolor=black,menucolor=black,runcolor=black,urlcolor=black]{hyperref}

\usepackage[normalem]{ulem}
\usepackage[usenames, dvipsnames]{color}
\usepackage{xcolor}
\usepackage{booktabs}
\usepackage{todonotes}
\usepackage{url}
\usepackage{algorithm}
\usepackage{algpseudocode}
\usepackage{amsmath}
\usepackage[nocompress]{cite}
\usepackage{url}
\usepackage{hyperref}
\usepackage{xcolor}
\usepackage{alltt}

\newcommand{\SuppOne}[0]{Supp.~\S1}
\newcommand{\SuppTwo}[0]{Supp.~\S2}
\newcommand{\SuppThree}[0]{Supp.~\S3}
\newcommand{\SuppFour}[0]{Supp.~\S4}
\newcommand{\SuppFive}[0]{Supp.~\S5}

\newcommand{\SuppExplainerBMatrix}[0]{Supp.~\S1.1}
\newcommand{\SuppExplainerGraphPrism}[0]{Supp.~\S1.2}
\newcommand{\SuppExplainerCensus}[0]{Supp.~\S2.2}
\newcommand{\SuppCensusImplementation}[0]{Supp.~\S2.1}
\newcommand{\SuppDataTable}[0]{Supp.~\S3.3}
\newcommand{\SuppDodecahedron}[0]{Supp.~\S3.4}
\newcommand{\SuppVisImplementation}[0]{Supp.~\S4.1}
\newcommand{\SuppVisFeature}[0]{Supp.~\S4.2}
\newcommand{\SuppVisGrid}[0]{Supp.~\S4.3}
\newcommand{\SuppSensitivity}[0]{Supp.~\S5}

\definecolor{ChangedContentColor}{HTML}{009900} 

\definecolor{AddContentColor}{HTML}{C7372F} 

\definecolor{DeleteColor}{HTML}{808080} 
    {} 

\definecolor{MoveColor}{HTML}{6495ED} 
    {} 

\definecolor{FixColor}{HTML}{C7372F} 
    {} 

\newcommand{\delete}[1]{}
\newcommand{\change}[1]{{#1}}
\newcommand{\move}[1]{{#1}}
\newcommand{\add}[1]{{#1}}

\begin{document}

\title{The Census-Stub Graph Invariant Descriptor}

\author{Matt~I.B.~Oddo,
  Stephen~Kobourov,
  Tamara~Munzner 
  \IEEEcompsocitemizethanks{
    \IEEEcompsocthanksitem Matt I.B. Oddo and Tamara Munzner are with the Department of Computer Science, University of British Columbia. Email: \{bofarull,tmm\}@cs.ubc.ca 
    \IEEEcompsocthanksitem Stephen Kobourov is with the Department of Computer Science at the Technical University of Munich. Email: stephen.kobourov@tum.de \protect\\}
  \thanks{Manuscript received Month XX, 202X; revised XXX.}
}

\markboth{IEEE TRANSACTIONS ON VISUALIZATION AND COMPUTER GRAPHICS, ~Vol.~xx, No.~x, xxxx~xxxx}%
{Author \MakeLowercase{Oddo \textit{et al.}}: The Census-Stub Graph Invariant Descriptor}

\IEEEtitleabstractindextext{%
\begin{abstract}

\change{An 'invariant descriptor' captures meaningful structural features of networks, useful where traditional visualizations, like node-link views, face challenges like the 'hairball phenomenon' (inscrutable overlap of points and lines). Designing invariant descriptors involves balancing abstraction and information retention, as richer data summaries demand more storage and computational resources. Building on prior work, chiefly the BMatrix—a matrix descriptor visualized as the invariant 'network portrait' heatmap—we introduce BFS-Census, a new algorithm computing our Census data structures: Census-Node, Census-Edge, and Census-Stub. Our experiments show Census-Stub, which focuses on 'stubs' (half-edges), has orders of magnitude greater discerning power (ability to tell non-isomorphic graphs apart) than any other descriptor in this study, without a difficult trade-off: the substantial increase in resolution doesn’t come at a commensurate cost in storage space or computation power. We also present new visualizations—our Hop-Census} \add{polylines} \change{and Census-Census trajectories—and evaluate them using real-world graphs,} \add{including a sensitivity analysis that shows graph topology change maps to visual Census change.} \textbf{Availability:} Our Supplemental materials are available at \textit{\href{https://osf.io/nmzra/}{osf.io/nmzra}}

\end{abstract}

}

\maketitle

\IEEEdisplaynontitleabstractindextext

\IEEEraisesectionheading{\section{Introduction}\label{sec:introduction}}


\begin{figure*}[!t]
\centering
\vspace{5pt}
\includegraphics[width=0.96\linewidth]{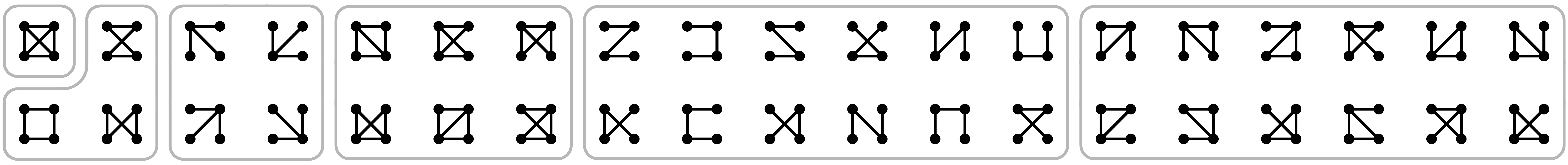}
\vspace{-4pt}
\caption{\add{Node-link views of all 38 labeled graphs with 4 nodes. With an unlabeled perspective there are only 6 graphs (groups by isomorphism class).}}
\vspace{-4pt}
\label{LabeledGraphs}
\end{figure*}


\IEEEPARstart{N}{etwork}~\change{visualization is increasingly needed in fields that rely on graph abstractions, such as chemistry, biology, medicine, finance, engineering, and transportation~\cite{WaterlooSurvey}.} The most intuitive representation of networks is the node-link idiom, where nodes are visually encoded as points in space and edges as lines that connect them~\cite{Bipartite_2022}.

\change{However, there are two challenges with the node-link idiom. The first challenge is the layout problem, where algorithms aim to reveal network structure saliently through different approaches, which can lead to the same network looking vastly different across representations~\cite{Kobourov_2013}. The second challenge is the hairball phenomenon, which happens with any network of large enough size and complexity, leading to the practical outcome of different networks all looking the same: a cluttered and overplotted mass of inscrutable points and lines~\cite{Munzner_2014}. Therefore, there is a need for alternative visual tools that can complement these two shortcomings. Graph descriptor data structures can provide a solution by summarizing network structure without relying on layout-dependent metrics—making the visualization of such descriptors particularly useful in complex scenarios, such as large graphs prone to layout hyper-variability and the hairball phenomenon.}

\change{To achieve visualizations impervious to traditional hurdles, we compute graph invariant descriptors, which are data structures that remain constant, no matter how network layout is manipulated~\cite{Tantardini_2019}.} \add{Descriptors are data summaries: trade-offs by definition, with the goal to be information-rich yet with a necessary degree of information loss~\cite{Bagrow_2024}.} \change{We design our invariant descriptor computation to carefully manage information loss, so that our summaries still capture structural meaning~\cite{Shvydun_2023}.} \add{Such invariant data summaries can be used to compare networks by anchoring network structure into a shared frame of reference.}

\change{We introduce the BFS-Census algorithm, a generalization and expansion of the classic Breadth-First Search (BFS) algorithm~\cite{Newman_2018}, that simultaneously counts three elementary constituents of network structure: nodes, edges, and stubs (half-edges). Accordingly, BFS-Census computes three Census data structures: Census-Node, Census-Edge, and Census-Stub—our three new invariant descriptors.}

\change{Since these three Census instantiations differ in degree of information loss, we introduce the Graph Atlas Collider quantitative analysis tool to evaluate how effectively invariant descriptors distinguish between non-isomorphic graphs. Our Collider tool leverages the Graph Atlas benchmark (the complete enumeration of undirected graphs from 3 to 10 nodes, a dataset of almost 12 million entries~\cite{Chen_2019}) to compute invariant descriptors from and tally their comparison results, including storage in bytes.}

Previous work on visualizing invariant descriptors has focused on heatmap visual encodings showing aggregated data: the BMatrix network portrait~\cite{Bagrow_2008} and its expansion introduced by the GraphPrism system~\cite{Kairam_2012}. We propose two new visual idioms that use line visual encodings to leverage the high discerning power of Census data structures:~\add{Hop-Census plots that are like parallel coordinates with an intrinsic order for the axes,} \change{and Census-Census plots that surface differential trajectory patterns.}

\change{To evaluate our new visual encodings with graphs larger than those in the Graph Atlas dataset, we curated a benchmark collection of 81 real-world and generated networks, chosen to represent a wide range of topology classes. We present a qualitative evaluation of invariant descriptor visualizations through analysis of static images generated from our visual encoding pipeline, which renders: our Census plots, visualizations from previous work (BMatrix network portrait), and the traditional node-link view.} \add{We also provide a sensitivity analysis that examines how degrees of topological change affect visual Census change.}

\change{In summary, in our five contributions we: \textbf{(1)}  propose the novel Census-Stub invariant descriptor, the algorithm for its computation, and an implementation; \textbf{(2)} present a quantitative discerning power evaluation of invariant descriptors, from our Graph Atlas Collider testing tool; \textbf{(3)} propose two new visual idioms, Hop-Census} \add{polylines} \change{and Census-Census trajectories, along with their implementations; \textbf{(4)} present a qualitative evaluation from our static image visual encoding pipeline, run on a diverse collection of 81 generated and real-world networks;} \add{and \textbf{(5)} present a preliminary sensitivity analysis, which shows significant topological changes result in substantial visual Census change, while minor changes yield correspondingly smaller effects.}

To support replicable research, we provide the Python code, data, and metadata as Supplemental materials.

\section{Motivation}

\add{First we show how the unlabeled perspective helps us focus on graph topology,} \change{then we explore graph descriptors, and finally we define invariant descriptor visualization.}

\vspace{-0.15cm}
\subsection{\add{Labeled vs Unlabeled Perspectives}}

\change{The topological structure of a graph is commonly defined by how its nodes are connected by edges.} \add{However, most previous approaches to graph drawing implicitly assume that graphs have meaningful labels attached to their nodes; we refer to this as the 'labeled graph perspective,' where identifying data exists independently of a node’s position within the network structure.}

\add{In contrast, the 'unlabeled graph perspective' identifies nodes solely by their positions within the network. For example, consider a graph of 4 nodes affixed to the corners of a square, as shown in Fig~\ref{LabeledGraphs}. How many unique ways can these nodes be connected? From the labeled perspective, the answer is 38. But from the unlabeled perspective, there are only 6 distinct ways to connect 4 unlabeled nodes, illustrated by the 6 groupings in Fig.~\ref{LabeledGraphs}. We say that graphs within the same set are isomorphic to each other,} \change{meaning they share the same 'graph topology'\cite{Newman_2018}: namely, the skeletal structure that remains unchanged,} no matter how the graphs are twisted, tangled, or have their nodes relabeled.

\change{While the labeled perspective is ubiquitous in traditional graph tasks, such as adjacency (which of the nodes A, B, or C has highest degree?), common connection (how many nodes are neighbors of both A and B?), and connectivity (what is the length of the shortest path between A and B?)~\cite{Lee_2008},} visualizations of unlabeled graphs—which are the target of this paper—\change{offer a visual summary that does not depend on additional node label data.} \add{The unlabeled perspective makes it natural to ask graph topology questions~\cite{Shervashidze_2011}, such as equivalence (}\change{how structurally similar are these two graphs?}\add{), regularity (does the graph have a recurring subgraph motif?), and category (}\change{are these two graphs from the same class of graphs?}\add{) to complement scenarios where traditional idioms fall short, like when the hairball phenomenon takes hold of the node-link idiom.}

\vspace{-0.15cm}
\subsection{\change{Data Abstractions and Invariant Descriptors}}


\add{Network data summaries involve a trade-off: concise abstractions lead to information loss, while richer ones increase data structure size~\cite{Bagrow_2024}. Balancing this trade-off is crucial for the design of effective invariant descriptors.} 

\change{In general, descriptors are data abstractions, or mathematical characteristics, that distill properties of a network into a concise data structure~\cite{Chen_2019}.} Some descriptors are relatively simple and can be derived by counting, like $n$ (number of nodes) and $e$ (number of edges), node degree (number of immediate neighbor nodes a node has), degree distribution (frequency of node degrees), diameter (the longest among shortest paths), eccentricity (maximum longest path from a given node), density (ratio of edges to the total number of possible edges), and clustering coefficient (fraction of triangles)—while others may require sophisticated transformations, such as a graph’s spectral matrix (eigenvectors of the adjacency matrix’s Laplacian)~\cite{Newman_2018}. Descriptors come in various shapes and sizes: they can be integer ($n$, $e$, diameter), decimal (density, clustering coefficient), vector (degrees, eccentricity), and array (spectral matrix)—indeed, graph descriptors can be collected into any data structure.


\change{We define 'invariant descriptors' as descriptors that do not vary between graphs of the same topological structure, considered from the unlabeled perspective.} Invariant descriptors are powerful graph analysis tools: if two graphs have different invariant descriptors, then we know with certainty they are non-isomorphic graphs. However, invariant descriptors are not one-to-one fingerprints; the analytical affordances of an invariant descriptor are tied to what we call 'discerning power', or its ability to avoid collisions. A collision happens when two non-isomorphic graphs have the same invariant descriptor. Usually, invariant descriptors that require complex data structures have higher discerning power and fewer collisions than those with simple data structures. For example, the diameter has low discerning power because many different graphs collide with the same integer value; but fewer graphs collide with the higher-powered degree distribution, which requires a multi-value vector as data structure (we experimentally verified this example in Sec.~\ref{sec:invariant-eval}).~\change{The existence of a zero-collision, one-to-one invariant descriptor is unknown, but unlikely—such a descriptor would be a perfect fingerprint and solve the famous graph isomorphism problem~\cite{Grohe_2020}.}

\vspace{-0.15cm}
\subsection{\change{Invariant Descriptor Visualization}}



\change{Descriptors are essential components of traditional network visualization. For example, to improve the clarity, readability, and overall effectiveness of a node-link visual encoding, layout algorithms use descriptors to appraise the graph drawing process: rewarding outcomes that reduce the number of edge crossings, avoid acute angles between same-node edges, minimize overstretched edges, and balance the aspect ratio of the finalized drawing~\cite{Purchase_2002}.} One of the most popular node-link layout approaches is the well-known family of iterative force-directed algorithms~\cite{Kobourov_2013}, which aim to optimize all these crossing, angle, overstretching, and aspect ratio metrics simultaneously.

\change{However, these descriptors, which depend on the drawing process itself, are examples of variant rather than invariant descriptors. In contrast, in this paper we focus on invariant descriptors by exploring their computation, quantifying their discerning power, and encoding them with dedicated visual idioms. We define 'invariant descriptor visualization' as visual encodings that rely only on information from an invariant descriptor's data structure. Furthermore, we propose deriving a coordinate system from this data structure, giving invariant descriptor visualizations a built-in frame of reference; a very useful property.} 

\delete{The visual encodings of highly discerning invariant descriptors can support unlabeled graph comparison tasks, because invariant descriptors do not depend on having known labels assigned to individual nodes~\cite{Tantardini_2019}. Indeed, because of their absolute frame of reference affordance, invariant descriptor visualizations are impervious to many of the hyper-variability and hairball challenges of traditional graph drawing techniques.}


\section{\change{Related Work}}

First we discuss node-link and adjacency matrix, followed by examples of non-traditional techniques. Then, we discuss the techniques we specifically build upon: the BMatrix network portrait and the GraphPrism system.

\vspace{-0.15cm}
\subsection{\change{Traditional Network Visualization}}

\delete{While the comparison of graph structural hallmarks is a critical challenge in network analysis~\cite{Tantardini_2019}, it is out of the scope of this paper to design specific tasks for unlabeled graph visualization using invariant descriptors, or to design distance metrics for graph analysis comparison. Instead, our motivation is to share a first approach to how highly discerning invariant descriptors can be used to visually explore graph topology.}

\begin{figure*}[!t]
\centering
\vspace{-2pt}
\includegraphics[width=0.96\linewidth]{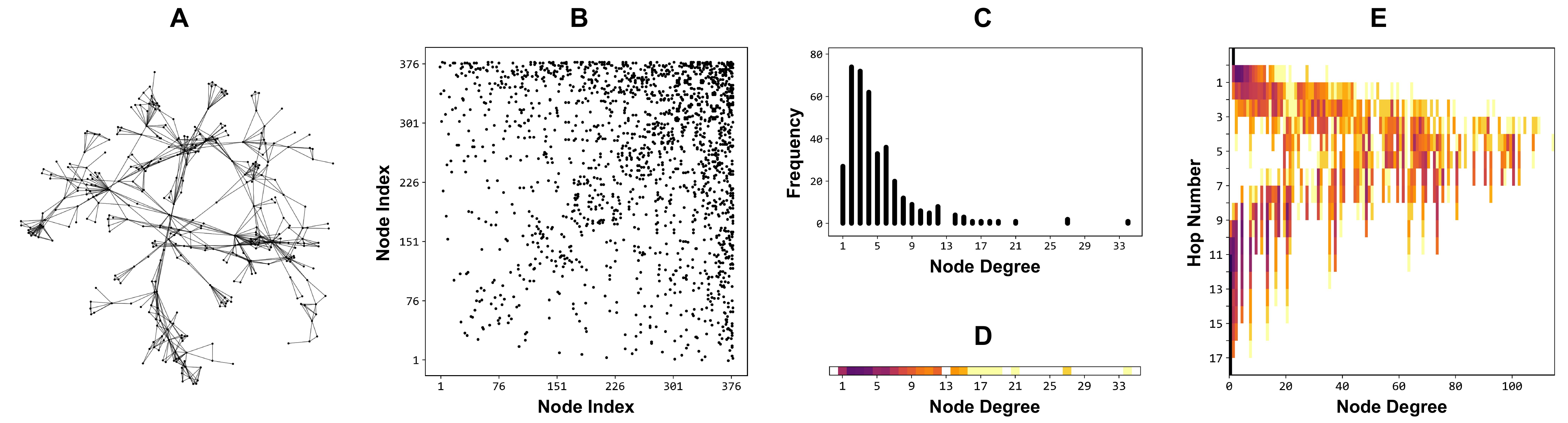}
\vspace{-6pt}
\caption{Multiple idioms representing the largest component within the \textit{dimacs10-netscience} network~\cite{NetworkScienceMain} from the \textit{konect.cc} repository, a graph we call \textit{network-science} in this paper. \textbf{(A)} The traditional node-link view, with force-directed placement. \textbf{(B)} Adjacency matrix, another traditional view, with nodes indexed randomly and then sorted by node degree.~\change{\textbf{(C)} Degree distribution plot, which is a histogram of the graph's node degree frequency. \textbf{(D)} Compact single-row heatmap of \textbf{C}, with frequency mapped to color. \textbf{(E)} The graph's network portrait~\cite{Bagrow_2008}, where node degree information is encoded in the horizontal axis, each degree's frequency of occurrence is color-encoded as a heatmap, and the vertical axis is a hop-comprehensive generalization of node degree accounting collected through BFS traversals. Note that \textbf{D} is equivalent to the 1-hop row in \textbf{E}.}}
\vspace{-5pt}
\label{IntroFigure}
\end{figure*}

\change{The node-link idiom (Fig.~\ref{IntroFigure}-A) is difficult to interpret for large or dense networks due to occlusion and clutter from overlapping lines and points: the so-called ``hairball problem''~\cite{Munzner_2014}. Even for smaller graphs, the upper limit of node-link layout effectiveness is a link density (number of edges divided by number of nodes) value of around 3, above which there is a surplus of edges that guarantee graph non-planarity~\cite{Melancon_2006}. Many alternative layouts have been proposed to mitigate hairballs, such as by optimizing metrics like node spacing in a geometric layout~\cite{McGuffin_2012}, making problematic edges invisible~\cite{Dianati_2016}, and minimizing the number and torsion of overlapping components and clusters~\cite{Archambault_2007}. However, hairballs inevitably emerge for large and complex networks despite these mitigations.} 

\change{Alongside the node-link idiom, the adjacency matrix (Fig.~\ref{IntroFigure}-B) is the other traditional network representation, which is robust against the hairball phenomenon and can better handle larger, dense networks.} The adjacency matrix is always a square data table, where rows and columns correspond to nodes, and non-zero cells indicate edges.~\change{An appropriate permutation of the node order in the rows and columns must be found for the adjacency matrix to show salient patterns, and even then tasks such as path tracing are poorly supported~\cite{Ghoniem_2004}.} This permutation problem is computationally challenging because there are $n!$ unique ways to permute a diagonally symmetric adjacency matrix of $n$ nodes, although many heuristic reordering algorithms have been proposed~\cite{McGuffin_2012}.~\change{Even though the data table abstraction fully describes a graph's topology, it does not do so uniquely, as many label permutations can represent the same graph. Data tables are not one-to-one invariant descriptors, because each permutation produces a distinct adjacency matrix visualization.}

\delete{Neither the adjacency matrix visualization nor its underlying data table are one-to-one graph invariant descriptors.}

\delete{The adjacency matrix idiom runs into scalability challenges when the number of matrix cells required exceeds available pixels, beyond roughly 1000 nodes for standard displays. This subpixel problem can be mitigated through interactive navigation (zoom and pan) and aggregation (row and column binning). However, these workarounds introduce legibility trade-offs.}

\vspace{-0.15cm}
\subsection{Alternative Network Visualization}

\change{We now discuss five non-traditional network visualizations that involve invariant descriptors in their design, although many use a mix of variant and invariant descriptors.}

\change{\textbf{Hive Plots}} are idioms where users arrange nodes on radial axes~\cite{HivePlot}, with line links between axes that can encode edges. \change{However, this technique also allows for cloned axes, in which case links that connect duplicate nodes behave like polylines between parallel coordinates~\cite{ParallelCoords}.} Therefore, Hive Plots are a family of radial network representations, which range from directly encoding nodes and edges as a version of the node-link idiom, to representations that aggregate away topological information in favor of computing dedicated descriptors for each axis, resulting in a version of parallel coordinate plots. The placement order within an axis; that is, the node permutation, can be computed according to any graph descriptor, variant or invariant~\cite{HivePlot2}. For example, a user can rank nodes in an axis by degree distribution, by clustering coefficient in another axis, and by eccentricity in another axis—all of which are permutations based on invariant descriptors.

\change{\textbf{Fabric Idiom}} is a technique that converts nodes to parallel horizontal lines (rows) and edges into parallel vertical lines (columns), where node-edge connections are encoded by dots at the perpendicular intersection of their respective lines~\cite{BioFabric}. \change{While this idiom avoids the direct overplotting problem of the hairball phenomenon, there is a problem with the expansive visual real estate taken by excess node and edge lines, which is driven by permutation trade-offs.}
In the BioFabric algorithm, columns can be permuted following ``cosine similarity'' or ``Jaccard similarity'', which are invariant descriptors.


\add{\textbf{Spectral Graph Drawing} is a family of graph-theoretic approaches that uses the eigenvectors of a graph’s Laplacian matrix (computed by subtracting the adjacency matrix from the diagonal matrix of node degrees), to embed node-link points in a two-dimensional space~\cite{Spectral2012}. While the computed eigenvalues are invariant descriptors and used to rank the eigenvectors, the spectral layout is still variant because node coordinates depend on eigenvectors, which are sensitive to adjacency matrix permutation.}


\change{Although all of these approaches do incorporate some form of invariant descriptor, they also introduce variant descriptors into the process, so the final visual encodings are ambiguous. In contrast to these previous approaches, we only use invariant descriptors.}

\change{\textbf{Graph Thumbnails} is a visualization system for exploring multiple complex networks simultaneously~\cite{GraphThumbnail}. Through $k$-core  decomposition,  nodes and edges of a given graph are aggregated into groups that are visualized as nested concentric bubbles. 
The hierarchical decomposition can be considered as an invariant descriptor. 
A related invariant descriptor, 'onion decomposition'~\cite{OnionS}, extends $k$-core to preserve node-specific information, enabling 'onion spectra' linecharts. The approach of $k$-core partitioning and aggregation has information loss leading to many collisions, whereas our approach provides much greater precision.}





\change{\textbf{Degree Distribution Plot} is a simple plot with node degrees on the horizontal axis and their frequency of occurrence on the vertical axis (histogram in Fig.~\ref{IntroFigure}-C, heatmap in Fig.~\ref{IntroFigure}-D). When sorted, the degree sequence data structure behind this plot is an invariant descriptor. Still, it has very low discerning power: information richness is limited to a 1-hop edge distance from any node. In contrast, we provide far more discerning power.}

\delete{A specialized version of the degree frequency plot appeared alongside the Watt-Strogatz and Barabási-Albert models of the late 1990s~\cite{Albert_2002}. These graph generator models were an important step towards modeling fully connected real-world graphs from social science: networks that feature a minority of highly connected nodes among a sparsely connected majority. When the degree frequency plot axes are log-transformed, the plot becomes a tool for evaluating power-law distribution, where points in log-log space are expected to follow a negative slope line. Despite visually encoding a simple data structure (a graph's degree sequence, which is a one-dimensional vector), the degree frequency plot, being an invariant descriptor visualization with an absolute frame of reference affordance, became a staple tool in the network science analytical toolkit.}


\vspace{-0.15cm}
\subsection{The BMatrix Network Portrait}
\label{sec:bmatrix}


\delete{An invariant descriptor visualization that captures more information through an algorithmic generalization of hop-based counting first appeared in 2008: a visualization technique called the network portrait~\cite{Bagrow_2008}.}

The network portrait (Fig.~\ref{IntroFigure}-E) is a heatmap encoding of the BMatrix data structure~\cite{Bagrow_2008} (also called matrix $B$~\cite{Bagrow_2019} or B-Matrix~\cite{Kairam_2012}), an invariant descriptor computed by what we call the BFS-BMatrix algorithm. Note that in this paper we name the algorithm (BFS-BMatrix) to be distinct from the data structure (BMatrix) and the visual encoding (network portrait) to facilitate visualization technique discussion~\cite{Nested_Model}.


\change{The BFS-BMatrix algorithm takes in a graph $G$ and generates a matrix $B$.} A matrix cell located by the coordinates $B[i][j]$ contains the frequency of occurrence of source nodes (an integer value) in the graph $G$ that have $j$ nodes in their $i$-th shell~\cite{Bagrow_2008}. A shell here refers to a set of nodes that are at the same shortest-path distance, or depth level (measured in number of hops), from the BFS source node. BFS-BMatrix works by iterating over each node in the graph, calculating the shortest path distances from that node to all other nodes, and then aggregates this information into a distribution that is then added to the matrix $B$, the BMatrix. The network portrait is the heatmap encoding of the integer values in the BMatrix cells. \change{In~\SuppExplainerBMatrix~we provide an illustrated algorithm walkthrough and complexity analysis.}


\delete{The time complexity of the well-known BFS algorithm is $O(n+e)$~\cite{Newman_2018}, where the best case topology is the tree graph, which has $n-1$ edges $e$, while the worst case is the complete graph, where $e$ is $n(n-1)/2$, simplified as $O(n^2)$ because each node $n$ is always connected to $n$ other nodes. The BFS-BMatrix algorithm compounds this complexity because it runs BFS for each source node $n$, so the tree graph scenario for BFS-BMatrix is $O(n^2)$, and $O(n^3)$ for the complete graph scenario. The computational complexity of BFS-BMatrix is proportional to density~\cite{Shvydun_2023}, and since most real-world graphs are closer in density to sparse trees than complete graphs, we expect BFS-BMatrix to perform at $O(n^2)$ in practice. Furthermore, the battery of BFS traversals that compute the BMatrix data structure can be parallelized~\cite{Bagrow_2008} by spreading the source nodes over multiple processors on a given machine~\cite{Czech_2017}.}


As a visual encoding, the network portrait presents an overview of connectivity that goes beyond the immediate neighbor relationships. \change{The first row captures information at 1-hop distance, which is equivalent to the degree frequency plot (Fig.~\ref{IntroFigure}-D). The key novelty of the BMatrix is in its subsequent rows, which compute the degree distributions of node degrees at incremental hops from the source node, making the BMatrix's height equivalent to the graph's diameter.}

\delete{To illustrate, the second row of the heatmap shown in Fig.~\ref{Fig02} is the node degree frequency at 2 hops away, then the third row is 3 hops away, then 4 hops, and so on.} 


\delete{Given that BMatrix is an invariant descriptor, its absolute frame of reference affordance makes the network portrait visualization impervious to many of the challenges traditional network visual encodings face. Moreover, BMatrix avoids scalability problems because matrix size depends on the graph diameter, which is typically much less than the total number of nodes.}

\change{While there is a visual resemblance to the adjacency matrix at first glance, it is important to recognize that the network portrait visually encodes the BMatrix—an aggregation data structure, from which the full graph topology cannot be reliably reconstructed. However, this is by design: the appeal of the BMatrix, as a graph summary, is that its matrix axes and heatmap patterns remain consistent across all isomorphs of a graph, unaffected by label permutation.}


\change{There were two key innovations of the BFS-BMatrix approach.} \add{First, using vectors to store traversal information in parallel with the classic BFS queue data structure, because the queue facilitates graph traversal but self-erases contents after nodes are visited.} \change{Second, generalizing node degree to capture multi-hop information rather than bound to a 1-hop limit. We preserve these key features in our own work while avoiding the information loss of matrix aggregation, and we also propose entirely new visual encodings of our invariant descriptors rather than simply using heatmaps.}

\delete{In this paper, we present the BFS-Census algorithm, which retains the key generalization of BFS that BFS-BMatrix innovates: employing vectors to collect and store traversal information in parallel with the classic BFS queue data structure—which facilitates graph traversal but self-erases queue contents after nodes are visited.}

\delete{The main contribution of the BMatrix is that it allows a hop-comprehensive generalization of the degree frequency plot, which is then encoded as a compact visualization of generalized node degree information~\cite{Bagrow_2019}.}

\vspace{-0.15cm}
\subsection{The GraphPrism System}


\change{GraphPrism~\cite{Kairam_2012} is a multi-faceted diptych that juxtaposes a force-directed node-link view against a vertical stack of different network portraits computed from the same graph.}

The algorithmic novelty of GraphPrism is the expansion of BFS-BMatrix to collect not just node degree information but five other invariant descriptors (respective facet names in parenthesis): node degree (Connectivity), density (Density), node clustering coefficient (Transitivity), node conductivity (Conductance), edge similarity (Jaccard), and edge ratio (MeetMin). These six facets share the same BMatrix data structure, guaranteeing that each matrix facet has the same number of rows. However, the number of columns differs across each matrix because the length of the BMatrix horizontal axis depends on the invariant descriptor collected. GraphPrism solves this horizontal axis alignment problem in two mitigation steps: first by cumulative computation using subgraphs, followed by binning to keep the number of columns consistent across facets. \move{In~\SuppExplainerGraphPrism~we explain these two steps with an illustrated example.}

\delete{First, GraphPrism introduces a cumulative version of the BFS-BMatrix algorithm, where each hop computes the invariant descriptor for the whole subgraph that is 1 hop away from the source node, then the subgraph 2 hops away, then 3 hops, and so on. Fig.~\ref{Fig03}-A shows the result, the cumulative version of the network portrait in Fig.~\ref{Fig02}. However, the horizontal axis can be large, leading to the subpixel problem. To address this problem—i.e. make the number of columns consistent across facets, and also ensure that matrix cells have a square aspect ratio—GraphPrism bins multiple matrix columns together so that the facets have the same data structure, shape, and size. Fig.~\ref{Fig03}-B shows the result, reproducing GraphPrism’s node degree (Connectivity) facet~\cite{Kairam_2012}.}

\change{We were inspired by the GraphPrism approach of expanding BMatrix to work with other descriptors, and we carry this idea further by proposing new invariant descriptors specifically designed to maximize discerning power. We designed our Census data structure to not only avoid aggregation (BMatrix) and cell binning (GraphPrism) information loss, but to further surface discerning patterns through the polyline visual encoding of Census vectors.}


\section{Algorithm and Data Structure}

We first provide an overview of our BFS-Census algorithm and Census data structure, then discuss both in more detail, and finally compare the Census and BMatrix data structures.

\vspace{-0.15cm}
\subsection{Overview}
\label{sec:alg-overview}



\change{We propose a new family of data structures, the Census invariant descriptors, which capture information about elementary constituents of graph topology: nodes, edges, and stubs. Nodes are points where edges start and end. Stubs, or half-edges, are the building blocks of an edge that always come in paired complements, where each stub is outbound from its respective node. In designing Census we aim to retain the strengths of BMatrix and its network portrait visual encoding, while also preserving node-specific information lost by aggregation during the computation of BMatrix. To achieve this, Census follows a 'bag-of-vectors' model}\add{—an unordered collection of vectors where each vector can vary in length, and repetitions are allowed.}


\change{When computed from the same graph, all Census instantiations share the same shape and size.} The number of Census vectors (corresponding to BMatrix rows) comes from the number of source nodes in $G$, but the lengths of Census vectors (corresponding to BMatrix columns) are determined by node degree computations during BFS traversal. This uniformity is because the shape and size of the bag-of-vectors data structure is fully determined by only one type of elementary constituent: nodes. When BFS reaches a node degree value of zero, the underlying BFS algorithm terminates, because it ran out of nodes to visit.

\change{The accounting of node degrees is an intrinsic component of BFS, so Census-Node can be computed by node degree counts alone.} However, the computation of any other Census instantiation still depends on BFS, and therefore also on node degree.~\change{Thus, Census-Edge and Census-Stub (and potentially other Census instantiations) inherit their data structure shape and size from Census-Node.}

\delete{Accordingly, we name the resulting instantiations Census-Node, Census-Edge, and Census-Stub.}

\begin{figure}[!htbp]
\centering
\includegraphics[width=0.98\linewidth]{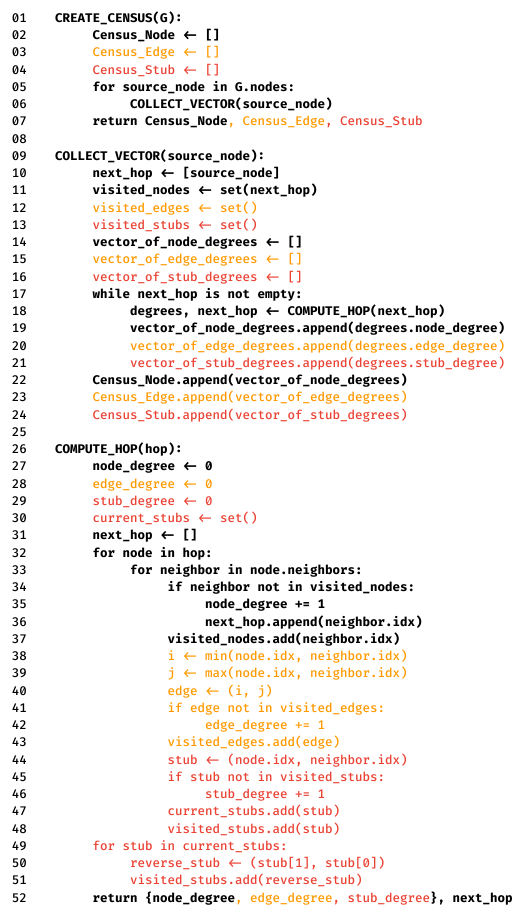}
\vspace{-7pt}
\caption{BFS-Census algorithm pseudocode. The concurrent computation of three Census instantiations is differentiated by color: Census-Node in black, Census-Edge in orange, and Census-Stub in red. Both Census-Edge and Census-Stub depend on Census-Node, but not vice versa.}
\vspace{-8pt}
\label{Algorithm}
\end{figure}

\begin{figure*}[!t]
\centering
\vspace{2pt}
\includegraphics[width=0.98\linewidth]{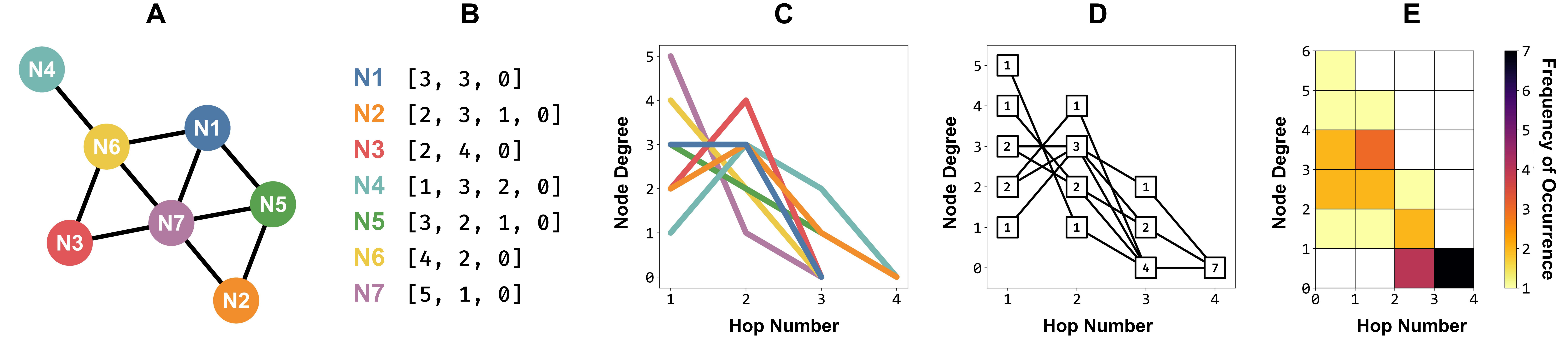}
\vspace{-5pt}
\caption{\change{Comparison of Census and BMatrix data structures and visual encodings regarding node-specific information loss. \textbf{(A)} Node-link layout of the input graph, with (arbitrary) node labels color-coded.} \textbf{(B)} Census data structure (Census-Node) computed from node degrees collection through BFS-Census. \textbf{(C)} Census vectors (invariant descriptors) are visually encoded as node-colored lines that encode node-specific information. \textbf{(D)} Aggregation into a matrix structure, counting frequencies of trajectories meeting at each grid cell, results in the graph’s BMatrix; node-based information is lost. \textbf{(E)} Matrix contents (frequencies) color-encoded as heatmap; node-based information remains lost.}
\vspace{-6pt}
\label{DataStructure}
\end{figure*}

\vspace{-0.15cm}
\subsection{Algorithm Details}
\label{sec:algorithm}


We present the full BFS-Census algorithm, which consists of three nested loops. The pseudocode (Fig.~\ref{Algorithm}) is color-coded by elementary constituent and Census instantiation being computed: Census-Node in black, Census-Edge in orange, and Census-Stub in red. \move{In~\SuppExplainerCensus~we provide an illustrated algorithm walkthrough and complexity analysis.}


The outermost loop is {\scriptsize {\sffamily {CREATE\_CENSUS}}} (Fig.~\ref{Algorithm} Line 01), which initializes the Census instantiations as empty lists (three bag-of-vectors), and starts the loop that iterates over the source nodes in the graph $G$. The middle loop, which we call {\scriptsize {\sffamily {COLLECT\_VECTOR}}} (Fig.~\ref{Algorithm} Line 09), takes an input source node from the outer loop and initializes for this node an assortment of empty vectors, one per invariant descriptor. The information that populates a vector's values comes from the computation of its elementary constituent degree at each hop (either node, edge, or stub), which takes place within the inner {\scriptsize {\sffamily {COMPUTE\_HOP}}} loop. 


The innermost loop, {\scriptsize {\sffamily {COMPUTE\_HOP}}} (Fig.~\ref{Algorithm} Line 26), initializes the integer variables that store the results of invariant descriptor computation. Notice that the integer {\footnotesize {\sffamily {node\_degree}}} and the list {\footnotesize {\sffamily {next\_hop}}} update simultaneously (Fig.~\ref{Algorithm} Lines 35 and 36), during the BFS traversal. \add{In practice, the list {\footnotesize {\sffamily {next\_hop}}} behaves as the classic BFS queue data structure, which self-erases (Fig.~\ref{Algorithm} Line 31) as traversal progresses.}

Within this innermost loop there is logic for control flow management through four sets: {\footnotesize {\sffamily {visited\_nodes}}}, {\footnotesize {\sffamily {visited\_edges}}}, {\footnotesize {\sffamily {visited\_stubs}}}, and {\footnotesize {\sffamily {current\_stubs}}}. These sets (lists that do not allow duplicates) ensure that BFS does not backtrack and incorrectly compute degrees that include already traversed nodes, edges, and stubs. The output of {\scriptsize {\sffamily {COMPUTE\_HOP}}} are the {\footnotesize {\sffamily {node\_degree}}}, {\footnotesize {\sffamily {edge\_degree}}}, and {\footnotesize {\sffamily {stub\_degree}}} integers. Each source node in $G$ gets a vector per invariant descriptor.


These vectors are collected into their respective bag-of-vectors Census instantiations in the middle loop. The middle {\scriptsize {\sffamily {COLLECT\_VECTOR}}} loop terminates when the {\footnotesize {\sffamily {next\_hop}}} list is empty (Fig.~\ref{Algorithm} Line 17). The outermost loop terminates when the BFS traversal of the last source node reaches a node of degree zero. Finally, the algorithm returns the Census-Node, Census-Edge, and Census-Stub invariant descriptors of $G$.

\vspace{-0.15cm}
\subsection{Data Structure Comparison}
\label{sec:datacompare}

\delete{We illustrate data structure differences with the same example graph of seven nodes as in Fig.~\ref{AlgorithmWalkthrough}.}

\change{We provide a detailed comparison between the Census data structure generated by our BFS-Census algorithm and the aggregated BMatrix produced by the previous BFS-BMatrix algorithm, highlighting how our approach preserves node-specific information that is lost in the latter.}

\change{Consider a BFS traversal of the graph in Fig.~\ref{DataStructure}-A with N7 as the source node.} N7 is highly connected, so at first hop BFS encounters 5 immediate neighbors (N3, N6, N1, N5, and N2), which get marked as visited. At the second hop BFS reaches 1 node (N4), the only node at 2-hop length, and marks it as visited. There are no more nodes to traverse at this point, so BFS performs one last hop, a third hop, and records the termination condition: 0 nodes visited (node degree zero). The last row in Fig.~\ref{DataStructure}-B shows \textbf{\footnotesize{\texttt{[5,1,0]}}} as the vector of node degrees at 1-hop, 2-hop, and 3-hop length from the N7 source node.

The Census data structure is made by conducting a BFS traversal and collecting such a vector for each node in the graph, with the hop number as the vector index. Fig.~\ref{DataStructure}-B shows the results, with color-encoded labels matching the source nodes on the Fig.~\ref{DataStructure}-A node-link view. Crucially, these vectors have different lengths: the Census data structure is a bag-of-vectors, not a rectangular matrix.

We plot the Fig.~\ref{DataStructure}-B Census vectors as lines in Fig.~\ref{DataStructure}-C, and keep their node coloring. The horizontal axis of Fig.~\ref{DataStructure}-C corresponds to the vector indices of Fig.~\ref{DataStructure}-B: these indices the number of hops away from the source node for each of the BFS traversals. The vertical axis corresponds to the values stored within the vectors: these numbers are generalized node degrees, or the number of new nodes encountered at each hop during BFS traversal.

Next, we can consider the Fig.~\ref{DataStructure}-C line plot as defining a rectangular grid with cells for each hop-degree combination, and sum the number of lines that pass through each hop-degree cell in the grid. Fig.~\ref{DataStructure}-D shows these aggregate counts. For example, 4 lines meet at the intersection of hop number 3 and node degree 0 in Fig.~\ref{DataStructure}-C (the lines of nodes N1, N3, N6, and N7). At the corresponding cell at the bottom next-to-right cell in Fig.~\ref{DataStructure}-D we see this aggregated value of 4 recorded as the frequency of occurrence.

While these 4 lines terminate at the 3rd hop in Fig.~\ref{DataStructure}-C, they must be extended into the 4th hop because the BMatrix expects all traversals to have the same length. Otherwise, each hop level in the aggregated matrix would not add up to the graph's total number of nodes (7 in this case). Finally, we can show these counts through color-encoding with a heatmap, as in Fig.~\ref{DataStructure}-E. This heatmap is exactly this graph's BMatrix network portrait, but with axes transposed when compared to Fig.~\ref{IntroFigure}-E.

\delete{In all subsequent network portrait views in this paper, we show BMatrix with this transposition, to facilitate comparison.}

\vspace{-0.15cm}
\subsection{\change{Census Data Structure Properties}}

The Census data structure contains all the information in BMatrix, and more. Unlike BMatrix, which aggregates vectors (Fig.~\ref{DataStructure}-D/E), Census retains node-specific information (node colors and labels in Fig.~\ref{DataStructure}-A/B/C). This means that we can
compute BMatrix from Census, but not vice versa.

Although the Census data structure shown in Fig.~\ref{DataStructure}-B includes node labels to make the walkthrough understandable, these labels are arbitrary and could be randomly permuted without affecting Fig.~\ref{DataStructure}-C/D/E. While Census does not rely on node labels, it can preserve them: BFS-Census directly passes along node index order from the input graph into Census vector order.

Census vectors are invariant descriptors; their values remain the same because values depend on the node's position relative to overall graph topology. The length of each vector corresponds to the node’s eccentricity, with the longest and shortest Census vectors reflecting the graph’s diameter and radius, respectively. Indeed, regardless of how nodes N2, N5, and N4 are labeled in different permutations, their Census vectors will always be the longest since these nodes are endpoints of this graph's diameter.


\section{\change{Data Abstraction Results and \\ Evaluation: The Graph Atlas Collider}}
\label{sec:invariant-eval}

We now describe and show results from our testing tool that quantifies invariant descriptor discerning power and storage size. \change{In~\SuppCensusImplementation~we detail our implementation.}

\change{Our Graph Atlas Collider evaluation tool gets its name from the Graph Atlas benchmark, an existing dataset that fully enumerates all tiny graphs up to 10 nodes, provided by Chen et al.~\cite{Chen_2019}.} While the creators call it a dataset of ``low-order non-isomorphic graphs'', we borrow the 'Graph Atlas' name from the first such atlas, then up to 7 nodes, published in 1998~\cite{Read_1998}. Although these tiny graphs may seem to be a 'toy' dataset, this benchmark is useful because it is a 'whole universe' dataset—it contains every possible graph topology within the node count bounds.

\change{We focus exclusively on fully connected topology, filtering out disconnected graphs from the outset. In total, our filtered Graph Atlas contains 11,989,762 connected non-isomorphic graphs with 3 to 10 nodes. While these nearly 12 million entries represent a large dataset, it remains manageable and allows for tests within a comprehensive set of low-cardinality graphs, from which insights can be applied to higher-cardinality real-world networks.} Moreover, because Graph Atlas is exhaustive, there
are no concerns about sampling bias or error.

\vspace{-0.15cm}
\subsection{Collider Architecture}

The design of our Collider tool is simple.~\change{Each invariant descriptor is handled independently, and we compute them for each graph entry in Graph Atlas. We partition Graph Atlas according to number of nodes; each partition called an 'order corpus'.} \add{From 3rd to 10th order, there are 2, 6, 21, 112, 853, 11117, 261080, and 11716571 graphs, respectively.} Graphs with different numbers of nodes cannot be isomorphic, so we do not check between different orders. Within each order corpus, our Collider performs a pairwise comparison with every other entry in that corpus. Whenever a pair of graphs share the same invariant descriptor, we count that pair as one collision for that descriptor.

\change{As output, a Collider test result datum is the total number of collisions per order corpus and per invariant descriptor. We define discerning power as the inverse of collision count; thus, one outcome of this quantitative evaluation is ranking descriptors based on their discerning power. Additionally, we analyze the trade-off between discerning power and storage space. Our Collider tool computes the bytes required by each descriptor, facilitating a comparison of data compression relative to discerning power.}

Specifically for storage analysis, we compute bytes from two descriptors to serve as baseline. One is the highly compact ’Graph6’ format~\cite{McKay_2000}, an ASCII string that can be decoded into an adjacency list; trading storage space for the computation time required to decode graph topology from it. \add{Graph Atlas is already stored in Graph6 format~\cite{Chen_2019}.} The other is the edgelist, a well-known format used by many node-link and adjacency matrix implementations, which we create to compute bytes from.

Notably, despite fully describing a graph's topology, neither the Graph6 nor the edgelist formats constitute one-to-one descriptors, because they depend on node label permutations—the same reason why the adjacency matrix data table is not an invariant descriptor.

\delete{Graph descriptors can also be considered as strategies for compressing topological information into a smaller data structure.}

\vspace{-0.15cm}
\subsection{\change{Experimental Design: Combinatorial Study}}

\begin{figure*}[!t]
\centering
\includegraphics[width=0.98\linewidth]{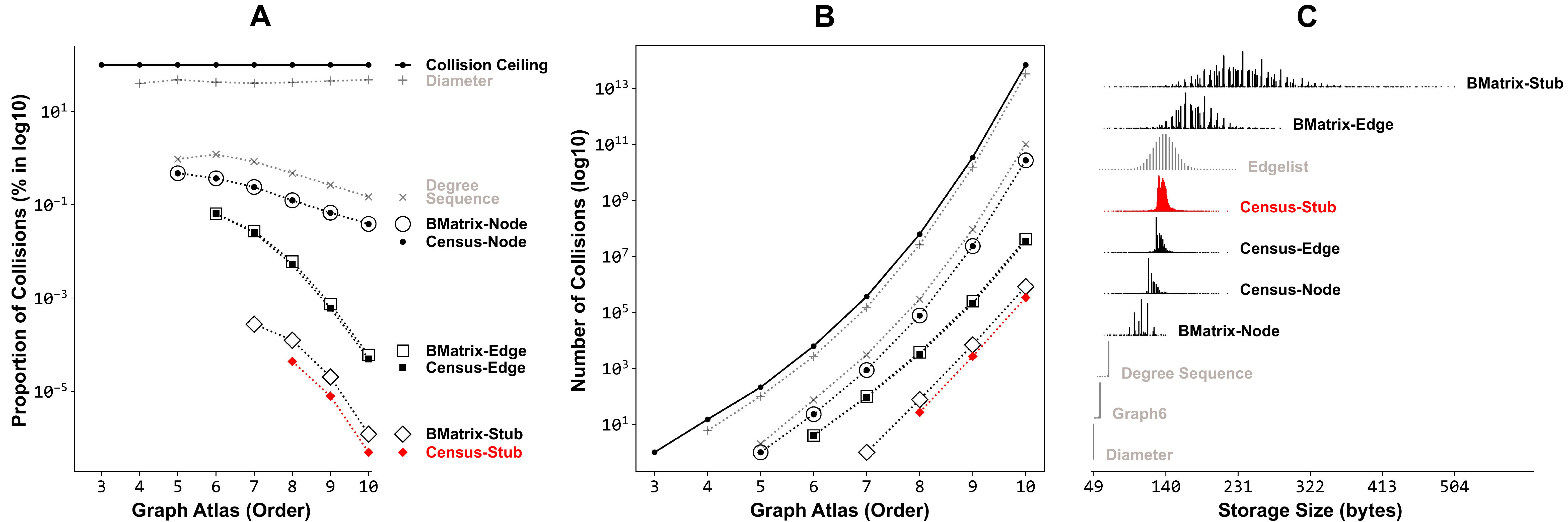}
\vspace{-4pt}
\caption{\change{Discerning power and storage space results. \textbf{(A)} Collision counts (log10) proportional to the maximum number of collisions per Graph Atlas order. The lower a line is, the better. Results show Census-Stub (shown in red) has orders of magnitude more discerning power than the other descriptors, that the choice of constituent matters the most, and that Census beats BMatrix in all cases. \textbf{(B)} Absolute numbers (log10) rather than proportions. \textbf{(C)} Histograms of invariant descriptor storage space, frequency normalized, each computed from the entire Graph Atlas. Sorted vertically by maximum bytes reached: the lower and narrower the histogram, the better. Unlike the exponential differences in discerning power in \textbf{A} and \textbf{B}, the differences across bytes are linear. While BMatrix-Stub is the most expensive descriptor in terms of storage size, Census-Stub is not.}}
\vspace{-6pt}
\label{ResultsCollision}
\end{figure*}

\begin{figure*}[!t]
\centering
\vspace{3pt}
\includegraphics[width=0.98\linewidth]{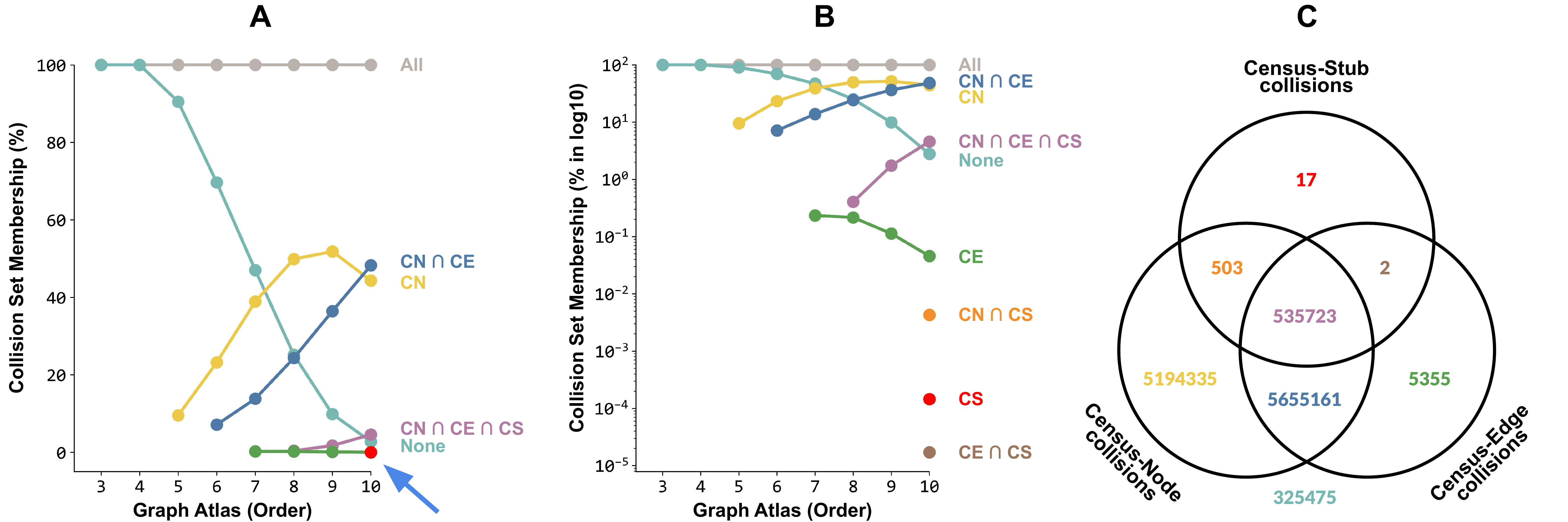}
\vspace{-4pt}
\caption{\add{Graphs can experience multiple, simultaneous collisions with their Census instantiations, allowing them to belong to one of eight intersecting sets. \textbf{(A)} Proportion of set membership across Graph Atlas to show large-scale patterns; the linear vertical axis leads to overplotting of the most-discerning sets (blue arrow). \textbf{(B)} A logarithmic vertical axis makes information visible, including that the 10th corpus is the only Graph Atlas order that includes graphs across all eight set intersections. \textbf{(C)} The 10th order corpus of 11,716,571 graphs with the counts of collision set membership.}}
\vspace{-6pt}
\label{ResultsSet}
\end{figure*}

\change{While the design of Graph Atlas Collider is straightforward, our choices about which invariant descriptors to compute for experimental analysis are more interesting.}

In order to quantify the range of differences in discerning power, we start by computing the maximum possible number of collisions, which we call the 'collision ceiling': \add{the theoretical upper limit of total pairwise comparisons for a Graph Atlas order.} The simplest case is the 3rd order corpus, where there are only two possible fully-connected topologies: the path graph (a line with 3 nodes and 2 edges) and the complete graph (a triangle of 3 nodes and 3 edges)—and therefore only one potential collision (line vs triangle). Each order corpus of $i$ entries has a collision ceiling $i\choose2$. In the 10th order corpus this number is 68.6 trillion collisions for the 11.7 million fully-connected graphs of 10 nodes. We conceptualize the collision ceiling as the result of an invariant descriptor with zero discerning power, because such a descriptor always guarantees a collision.

\delete{This number is the count of all possible non-isomorphic pairings within an order corpus.}

\delete{Conceptually, the opposite of the collision ceiling is the 'collision floor', which is theoretical: defined by a zero-collision perfect fingerprint invariant descriptor that solves graph isomorphism.}

We design our Collider testing as a combinatorial study, exploring the effects between what is topologically counted (node vs.~edge vs.~stub) and how these counts are collected (Census vs.~BMatrix). Our BFS-Census algorithm computes for each Graph Atlas entry its Census-Node, Census-Edge, and Census-Stub invariant descriptors. Then, through the process shown in Fig.~\ref{DataStructure}, we convert these Census instantiations into their respective BMatrix aggregations, which we call BMatrix-Node, BMatrix-Edge, and BMatrix-Stub. To check the integrity of BMatrix conversions, we also include the BMatrix computed by the original network portrait algorithm~\cite{Bagrow_2019}, which we call BFS-BMatrix. We expect the BMatrix from BFS-BMatrix to be identical to our Census-derived BMatrix-Node, from BFS-Census.

Outside of the combinatorial study, we also include two invariant descriptors with low discerning power that require small data structures: diameter, which is an integer; and degree sequence, which is a one-dimensional vector. We expect both diameter and degree sequence to be closer to the collision ceiling than any of the other descriptors in this experiment, and between these two, we expect diameter to be the least discerning, given it is only an integer.

\delete{We use the sorted degree sequence of a graph as the instantiation of degree distribution.}

\delete{As an additional comparison, we also compute two simpler data structures: the graph's diameter, and its degree sequence.}

\delete{These amount to 6 invariant descriptors, of which 3 are the Census instantiations Census-Node, Census-Edge, and Census-Stub, and the other 3 are their respective BMatrix-Node, BMatrix-Edge, and BMatrix-Stub aggregations.}

\delete{Finally, we also include an integrity validation check that compares the original BFS-BMatrix data structure~\cite{Bagrow_2008} against our BMatrix-Node derived from Census-Node.}

\vspace{-0.15cm}
\subsection{\change{Discerning Power Results}}

\change{We present the total count of collisions for each invariant descriptor (in logarithmic scale) across Graph Atlas orders in Fig.~\ref{ResultsCollision}-A/B. To facilitate comparison between orders and reduce line overlap, Fig.~\ref{ResultsCollision}-A shows the collision counts as a ratio of the collision ceiling at each order. In contrast, Fig.~\ref{ResultsCollision}-B displays raw collision counts without ratio conversion.}

\change{In Fig.~\ref{ResultsCollision}-A/B, a lower line position on the vertical axis indicates better performance for an invariant descriptor: a greater distance from the collision ceiling means higher discerning power. Our results show that Census-Stub has orders of magnitude more discerning power than the other invariant descriptors in this study. For the 10th order corpus of Graph Atlas in Fig.~\ref{ResultsCollision}-B, the range spans five orders of magnitude: the number of Census-Stub collisions is under 337 thousand, compared to 34 million for Census-Edge, and over 26 billion for BMatrix-Node.}

Regarding combinatorial results, we see that the most consequential choice is the elementary constituent counted. Stubs have the highest discerning power, followed by edges, with nodes having the least discerning power. We find that in all cases the Census data structure has higher discerning power than the BMatrix. The improvement in discerning power between the Census and BMatrix data structures is almost invisible in Fig.~\ref{ResultsCollision}-A/B for nodes and edges, but the superior power of the Census data structure is clearly visible in the case of stubs. 

The integrity check validates our claim that our BMatrix-Node (computed from Census-Node) is identical to the original BMatrix. \add{In~\SuppDataTable~we provide a table with all collision count results, including integrity validation.}

\delete{In Fig.~\ref{TGC_Results}, the collision count line of the traditional BMatrix (computed from the BFS-BMatrix algorithm, labeled BFS-BMatrix in Fig.~\ref{collisions_TGC}) is not explicitly shown because it is identical to the BMatrix-Node line and would directly overlap.}

\vspace{-0.15cm}
\subsection{\change{Storage Space Results}}

Our empirical storage results are shown in Fig.~\ref{ResultsCollision}-C. All invariant descriptors are stored as serialized strings, except for Graph6 which is already in string format. One notable result is that the storage requirements of all three of the Census instantiations are very similar to each other, in sharp contrast to the three BMatrix instantiations, where the storage requirement changes size substantially in response to the discerning power of the invariant descriptor collected.

The benefits of the Census data structure are clear: the very superior discerning power of Census-Stub is available for the same size cost of Census-Node and Census-Edge. However, BMatrix-Stub requires much more space than Census-Stub; and BMatrix-Edge requires somewhat more space than Census-Edge. The original BMatrix, or BMatrix-Node, is more parsimonious in size, requiring slightly less space than the Census variants. In terms of magnitude, the differences in compression are roughly linear: all descriptors are contained within a 49 to 504 byte range (Fig.~\ref{ResultsCollision}-C). 

While the differences in discerning power are many orders of magnitude, the differences in storage space are very small, roughly a factor of two. Overall, our results show that there is no difficult trade-off to navigate: the huge increase in Census-Stub discerning power do not come at a commensurate cost of storage space.

\vspace{-0.15cm}
\subsection{\add{Collision Set Membership Results}}

\add{Another result of our combinatorial study is that we can classify the multiple types of collisions that a single Graph Atlas entry can have, as shown in Fig.~\ref{ResultsSet}. We assign each entry in Graph Atlas a tag indicating any Census-Node, Census-Edge, or Census-Stub collision to explore how sets of Census collisions, and their intersections, are distributed within Graph Atlas. We focus only on Census intersecting sets and do not conduct this analysis for BMatrix instantiations, because Census is the more discerning data structure.}

\add{Across Graph Atlas orders, there are always graphs that experience no collisions. These collision-free graphs are abundant in the lower orders of Graph Atlas, but their presence rapidly declines as order increases, as shown by the \textit{None} set (teal line) in Fig.~\ref{ResultsSet}-A. Conversely, graphs that experience simultaneous collisions across all three Census instantiations are rare and only emerge starting with the 8th order, with an upward trend as order increases, as shown by the $\mathit{CN}\cap\mathit{CE}\cap\mathit{CS}$ set (purple line) in Fig.~\ref{ResultsSet}-B.}

\add{We refer to sets of graphs that have only one of Census-Node, Census-Edge, or Census-Stub collisions as a 'singleton' set. Interestingly, we see the Census-Node singleton (yellow \textit{CN} line in Fig.~\ref{ResultsSet}-A) has its maximum at the 9th corpus, where over 50\% of graphs have a Census-Node collision. In contrast, the Census-Edge and Census-Stub singletons have negligible presence: at the 9th corpus, the Census-Edge set (green line) has 0.1\% of the graphs, and the Census-Stub set has none, which means that data structure has full discerning power for graphs up to 9 nodes. Only in the 10th order corpus does the Census-Stub singleton appear (red point), composed of just 17 graphs, amounting to 0.0001\% of the 10-node graphs.}

\add{The rarest collision set is $\mathit{CE}\cap\mathit{CS}$ (brown point in Fig.~\ref{ResultsSet}-B), with only 2 graphs across the entirety of Graph Atlas experiencing this intersection set (Fig.~\ref{ResultsSet}-C). Additionally, the second and third rarest are the \textit{CS} singleton (red point) and $\mathit{CN}\cap\mathit{CS}$ intersection (orange point), respectively, both exclusive to the 10th corpus. The fourth rarest is the \textit{CE} singleton (green line), which emerges from the 7th order onwards. Finally, the fifth rarest, which can be seen in Fig.~\ref{ResultsSet}-A, is the $\mathit{CN}\cap\mathit{CE}\cap\mathit{CS}$ intersection (purple line), which shows that in almost all instances where Census-Stub has a collision, it co-occurs with other Census instantiations. In~\SuppDataTable~we provide a table of set membership results.}

\vspace{-0.15cm}
\subsection{\add{Collision Tuple Highlights}}

\begin{figure}[!htbp]
\vspace{2pt}
\centering
\includegraphics[width=0.86\linewidth]{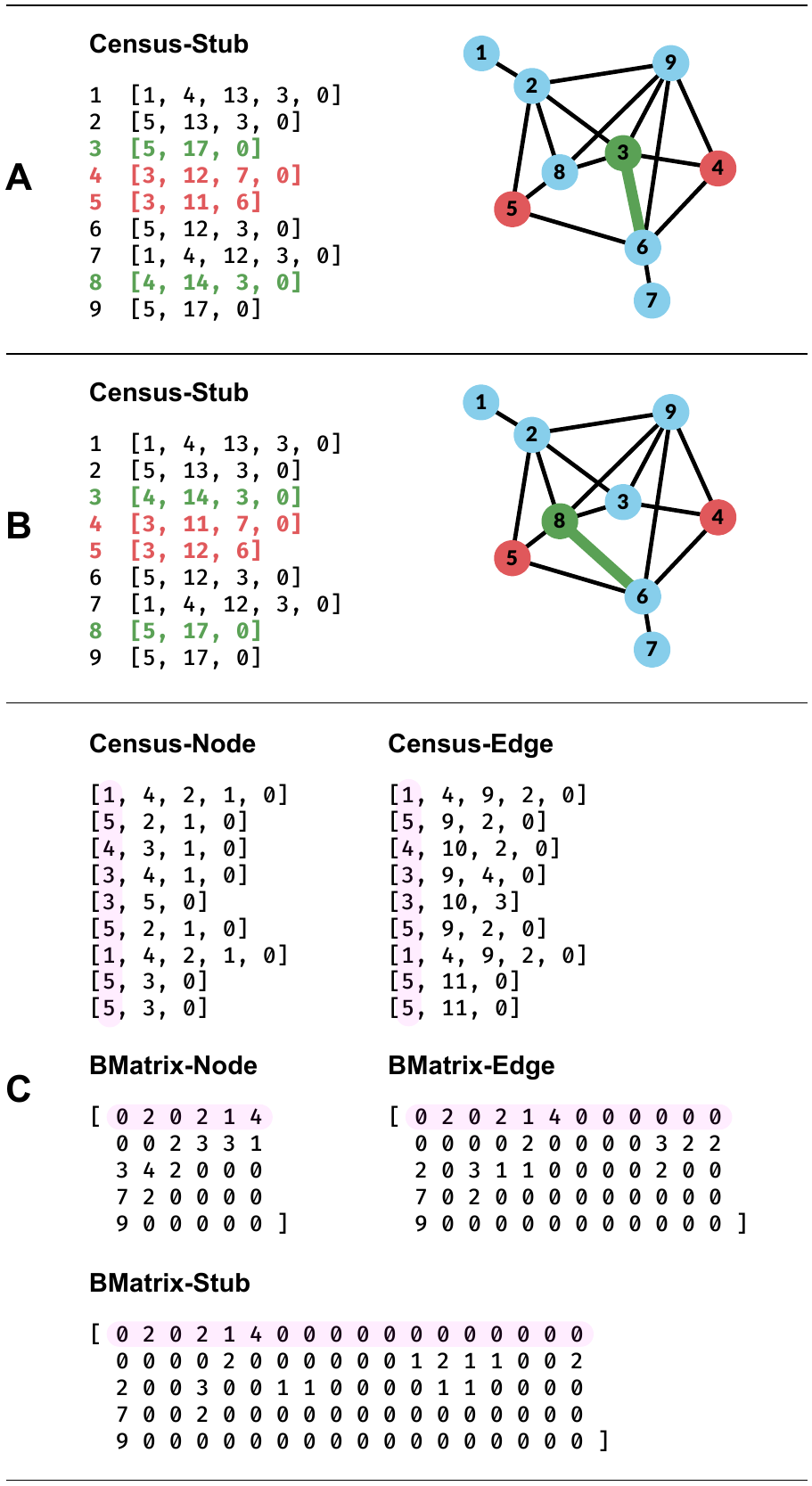}
\vspace{-3pt}
\caption{\add{Collision tuple and Census properties. Two non-isomorphic graphs, each with 9 nodes and 16 edges, are shown with Census data structures on the left and node-link views on the right. These can only be distinguished by Census-Stub; every other invariant descriptor results in a collision. \textbf{(A)} The first graph (Graph6: H?bFUiN) differs from the second only in the edge between nodes 6 and 3 (green). \textbf{(B)} The second graph (Graph6: H?H?bFSzF) has this edge between nodes 6 and 8, causing the locations for the Census vectors of nodes 3 and 8 to swap compared to the first graph (green) and, critically, altering the values in the Census vectors for nodes 4 and 5 (red). \textbf{(C)} The 5 other descriptors are identical for both graphs. The shared degree sequences are captured in the first values of the Census vectors (purple), and lead to the degree distributions in the first BMatrix rows (purple).}}
\vspace{-7pt}
\label{CollisionHighlight}
\end{figure}

\add{Our Collider tool identifies 'collision tuples': all of the non-isomorphic graphs that share identical invariant descriptors. In Fig.~\ref{CollisionHighlight} we present a collision 2-tuple (a pair) that only Census-Stub distinguishes, while all other invariant descriptors in this study yield collisions. These two 9-node graphs are in the $\mathit{CN}\cap\mathit{CE}$ set (blue line in Fig.~\ref{ResultsSet}-A/B).}

\add{Fig.~\ref{CollisionHighlight} also uses this collision pair to highlight several properties of the Census data structure by showing the full numeric values for all Census and B-Matrix invariant descriptor data structures. A key property of Census is that all vectors capture the same information at the first hop, so the first column of 1-hop values (purple columns in Fig.~\ref{CollisionHighlight}-C) is the graph's degree sequence. Thus, the number of rows (vectors) is the graph's number of nodes $n$, and the sum of the first column (1-hop values) divided by two is the number of edges $e$. When aggregated, this information becomes the degree distribution visible in the first row of the BMatrix (purple rows in Fig.~\ref{CollisionHighlight}-C), for example the value of $4$ in the final (index 5) position is the count of the number of $5$s that appear in the first column of Census.}

\add{Another data structure property is that the Area Under the Curve (AUC) of any Census-Node vector equals the graph's number of nodes minus one (since BFS-Census excludes the source node); in Fig.~\ref{CollisionHighlight}-C, we see that each row of the Census-Node data structures sums to 8.
Similarly, the AUC of a Census-Edge vector equals the total number of edges, and each of those rows sum to 16. Census-Stub is special: its AUC varies due to a blocked backtracking step in its computation (Sec.~\ref{sec:algorithm}; Fig.~\ref{Algorithm}, Lines 49–51), which causes differential stub accounting depending on source node position within network structure. In~\SuppExplainerCensus~we discuss this property in more detail.}

\move{In~\SuppDodecahedron~we highlight a collision noted in previous work: the network portrait (BMatrix-Node) collision of the Dodecahedron and Desargues graphs~\cite{Bagrow_2019}. We show Census-Stub can discern between these two 20-node graphs.}


\begin{figure*}[!t]
\centering
\vspace{2pt}
\includegraphics[width=0.99\linewidth]{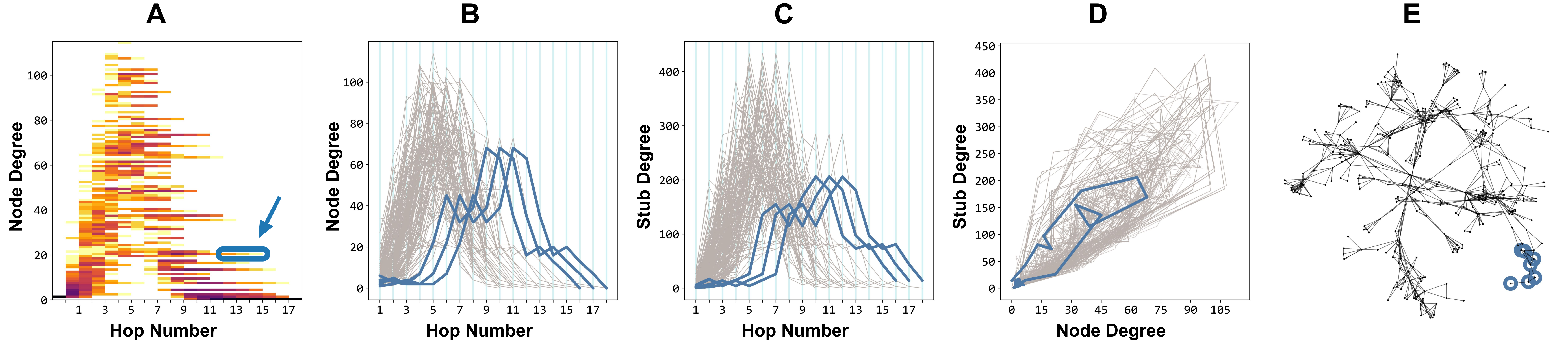}
\vspace{-3pt}
\caption{Visual encodings of invariant descriptors, computed from \textit{network-science} (the same network as in Fig.~\ref{IntroFigure}-A). \textbf{(A)} Heatmap Matrix plot of BMatrix-Node computed from Census-Node, a transposed version of the network portrait in Fig.~\ref{IntroFigure}-E, with three BMatrix cells selected (blue arrow and highlight). \textbf{(B)} Hop-Census plot of Census-Node, with node-specific information from Census revealing the identity of the selected nodes (blue \add{polylines}) that in \textbf{A} aggregate into the selected BMatrix cells. \textbf{(C)} Similarly, \add{polylines} of Census-Stub, which has the same shape and size as Census-Node, but with higher discerning power. \textbf{(D)} Census-Census plot of Census-Node perpendicular to Census-Stub (CN-CS plot), with highlighted trajectories in blue. \textbf{(E)} Node-link view, highlighted in blue are the seven source nodes from which the blue trajectories are from.}
\vspace{-6pt}
\label{EchoPattern}
\end{figure*}

\section{Visual Encodings}
\label{sec:visenc}

We now discuss approaches for visually encoding invariant descriptors. First, we revisit the previously proposed Heatmap Matrix. \change{Then, we introduce our new plots, which better exploit the high-precision information of Census: Hop-Census polylines for a single Census data structure, and Census-Census trajectories that capture differentials between two Census instantiations orthogonal to each other.}

\vspace{-0.15cm}
\subsection{\change{Aggregation:} Heatmap Matrix Plots}


Heatmaps have been the standard way to show the aggregate node information in the BMatrix data structure, visually encoding matrix cell values with a logarithmic colormap. It is straightforward to extend this visual idiom to represent information about edges (BMatrix-Edge) and stubs (BMatrix-Stub) as well. In contrast to the orientation of the network portrait in Sec.~\ref{sec:bmatrix} (Fig.~\ref{IntroFigure}-E), we will transpose Heatmap Matrix plots from this point onwards in the paper, as shown in Fig.~\ref{DataStructure}-E, to align respective axes with our Hop-Census visual encoding.

\delete{(\textit{inferno} from Python's \textit{matplotlib} library in this paper)}


In Fig.~\ref{EchoPattern}-A, there is a salient recurring pattern, where the horizontal regions on the right side appear to \change{'echo'} across; these also occur transposed in Fig.~\ref{IntroFigure}-E, as vertical downwards 'drips'. These patterns are described as ``fine-scale structure'' in the original network portrait paper~\cite{Bagrow_2008}, and linked to differences in network assortativity; that is, how tree-like the graph topology is, with more tree-like networks having more of these patterns. In Fig.~\ref{EchoPattern}-A, we have highlighted and selected one of these \change{'echo'} patterns: three BMatrix cells located at node degree 20 and hop 13, 14, and 15. These patterns cannot be explained through a heatmap matrix alone, because the underlying BMatrix lacks node-specific information. \change{To understand what these 'echo' patterns are, we must visualize Census vectors as polylines.}

\vspace{-0.15cm}
\subsection{\add{Polylines with Ordered Axes:} Hop-Census Plots}


We first consider the visual encoding of Census vectors as multiple \add{polylines}, all superimposed within the coordinate frame. We call this visualization a Hop-Census plot, of which Fig.~\ref{EchoPattern}-B/C and Fig.~\ref{DataStructure}-C in Sec.~\ref{sec:datacompare} are examples. 


The horizontal axis is the number of hops, just as in the heatmap view of Fig.~\ref{EchoPattern}-A. The vertical axis is the count of topological constituents for that Census instantiation: node degrees from the Census-Node in Fig.~\ref{EchoPattern}-B, and stub degrees from the Census-Stub in Fig.~\ref{EchoPattern}-C. \change{A Hop-Census plot can also be generated for Census-Edge. Unlike the Heatmap Matrix plot, which aggregates degree count information through the BMatrix, Hop-Census plots directly display the higher precision node-specific information within the Census data structure, encoded as} \add{polylines}.


\add{The 'polyline' (a connected, jagged set of straight line segments~\cite{Munzner_2014}) is the default line encoding of parallel coordinate plots, a visualization technique used to display multivariate data in which each axis represents a different variable, with polylines threading common points across these variables~\cite{ParallelCoords}. While a shortcoming of parallel coordinates is that axis ordering is arbitrary, in Hop-Census plots the axis order is strictly determined by hop number. Thus, while Hop-Census plots are visually similar to parallel coordinates, they do not share that shortcoming due to the characteristics of their underlying data abstraction.} 

\add{In Fig.~\ref{EchoPattern}-B/C, hops are directly encoded as vertical cyan bars, to highlight their interpretation as parallel coordinates. However, unlike traditional parallel coordinates, where each axis is normalized independently, our Hop-Census axes share the same vertical range of values, which are directly derived from the respective Census instantiation. Thus, we exploit the absolute frame of reference provided by the Census data structure to allow comparison between datasets to be meaningful with this visual encoding.}


We can leverage the fact that the axes of Fig.~\ref{EchoPattern}-A (BMatrix-Node) and Fig.~\ref{EchoPattern}-B (Census-Node) are the same to explain the \change{'echo'} pattern using the \add{polylines} encoded from Census-Node vectors. As shown numerically in Fig.~\ref{DataStructure}, the three BMatrix-Node cells selected in Fig.~\ref{EchoPattern}-A encode the number of \add{polylines} that pass through these cells. With the matrix coordinates of these cells, we can retrieve the respective \add{polylines} from Census-Node, which in Fig.~\ref{EchoPattern}-B are highlighted in blue. We find seven \add{polylines} from Census-Node vectors that aggregate into the three selected cells. These vectors come from seven different nodes, highlighted on the Fig.~\ref{EchoPattern}-E node-link diagram.


The \add{polylines} of these seven nodes reveal that the \change{'echo'} pattern is a repetition in invariant descriptor values, which can be interpreted as a lag propagating along Census hops (the cyan bars in Fig.~\ref{EchoPattern}-B/C). This lag occurs because the selected nodes are located in the periphery of the graph's topology (Fig.~\ref{EchoPattern}-E). To illustrate, when BFS-Census traversal begins with one of these peripheral nodes as source node, it requires a varying number of hops to reach the main body of the graph. Crucially, there exists a topological bottleneck that each of these traversals must go through before reaching the main body of the graph, after which point vector information becomes identical. However, the values are delayed by the number of hops already traversed to arrive at the bottleneck. \change{We conclude that the 'echo' pattern emerges from lag in polyline values that stem from nodes that share placements on the same side of a topological bottleneck.}

\vspace{-0.15cm}
\subsection{\add{Differential Trajectories:} Census-Census Plots}

The Census data structures encoded in Fig.~\ref{EchoPattern}-B (Census-Node) and Fig.~\ref{EchoPattern}-C (Census-Stub) have similar, but distinct \add{polyline} patterns. Their similarity arises from the fact that Census vectors are always well aligned, as all Census instantiations share the same data structure shape and size.

To highlight the differences between two Census instantiations, we can plot one Census data structure orthogonal to another. We refer to this visual encoding as a Census-Census plot, an example of which is shown in Fig.~\ref{EchoPattern}-D. Census-Census plots visually encode the information stream of two \add{polylines} as a single line; in Fig.~\ref{EchoPattern}-D, the information from Census-Node (horizontal axis) is perpendicular to Census-Stub (vertical axis). \add{In contrast to the Hop-Census plot, where the parallel axes encode the number of hops, the Census-Census plot encodes hop number through the number of points that make up a line. We refer to the line that threads these points as a 'trajectory', because unlike the equidistant horizontal span of polylines, trajectory points can land anywhere in Census-Census space.}

By plotting one Census instantiation perpendicular to another, both hop dimensions are encoded in a two-dimensional space, highlighting where trajectories are the same and where they differ. For example, in Fig.~\ref{EchoPattern}-D, the trajectories of the seven nodes no longer display the lag pattern, instead they converge into a single shared line: the seven trajectories eclipse each other. The only region in the Census-Census plot where this single trajectory is not shared is near the Fig.~\ref{EchoPattern}-D origin: the seven nodes differ only in their first hops before reaching the topological bottleneck.

Census-Census plots make visually salient patterns of differential information between the trajectories of two Census instantiations, whereas these differential patterns do not surface in Hop-Census plots.


\begin{figure*}[!tbp]
\centering
\vspace{2pt}
\includegraphics[width=0.98\linewidth]{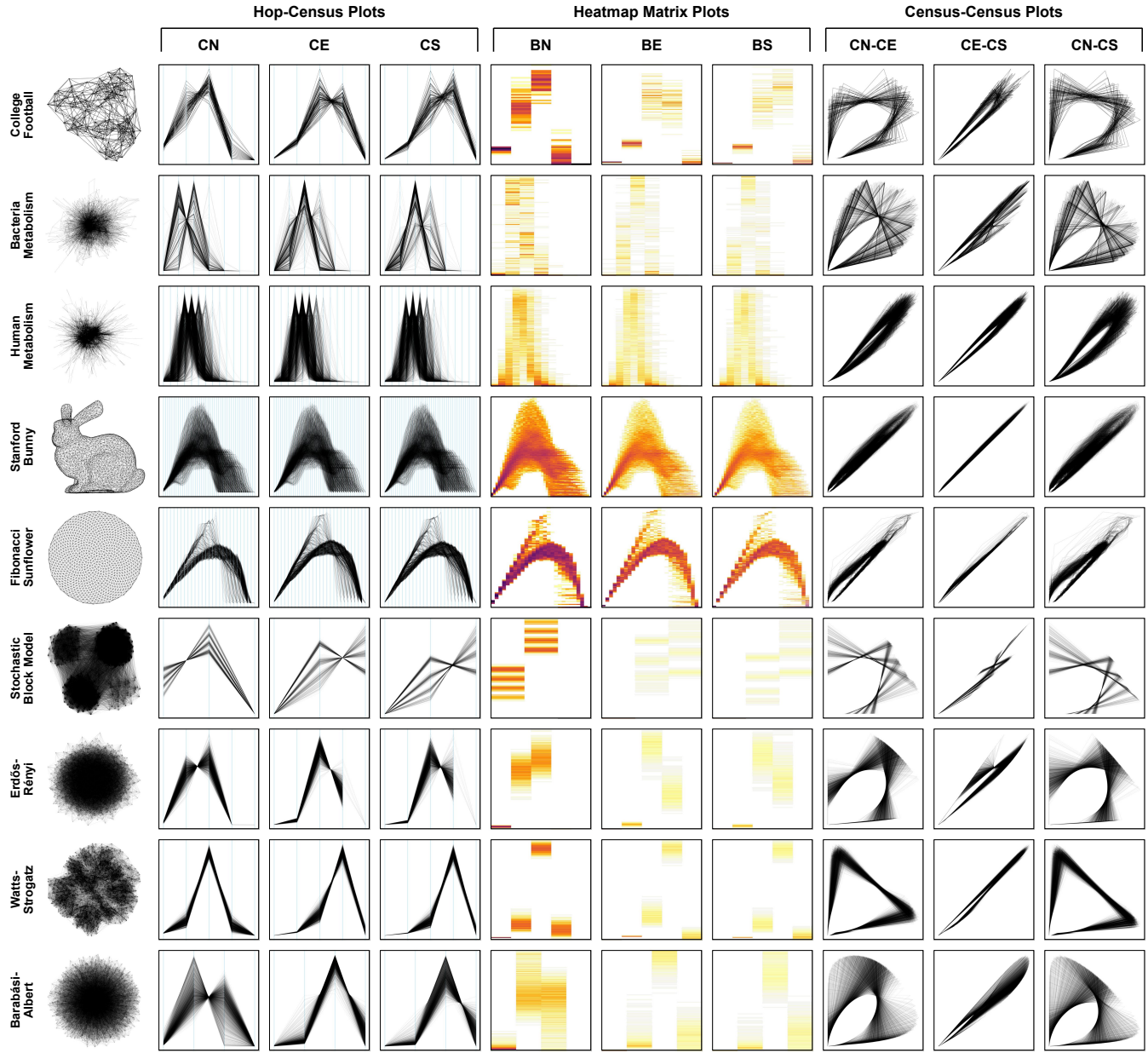}
\caption{Visual encodings for nine different networks, selected from our eclectic dataset of 81 real-world and generated networks assembled to show behavior across a wide variety of graph types. From left to right, Hop-Census \add{polyline} plots for each of the Census instantiations; followed by the Heatmap Matrix plots of their respective BMatrix aggregations; and the three possible Census-Census trajectory plot combinations of Census-Node (CN), Census-Edge (CE), and Census-Stub (CS). \change{All plots normalized within the unit square;~\SuppVisFeature~contains full non-normalized versions.}}
\vspace{-6pt}
\label{SelectedResults}
\end{figure*}

\section{\change{Visual Encoding Results and Evaluation}}

We first discuss the architecture of our visual encoding pipeline, then summarize the datasets that we explore using it. We then present our qualitative analysis of the resulting static images, and discuss the affordances of the absolute frame of reference that our approaches exploit.

\vspace{-0.15cm}
\subsection{\change{Encoding Pipeline Architecture}}
\label{sec:encodearch}

\change{Our Python-based visual encoding pipeline renders 10 static images for each input graph. It uses our implementation of the six Census plots that we propose: three Hop-Census} \add{polyline} \change{plots and three Census-Census trajectory plots.} It also creates heatmap matrix plots for each of the three constituents (nodes, edges, and stubs) using the \textit{inferno} colormap, and a node-link view of the input graph. \add{Slight variations of this rendering pipeline were used to create all figures in this paper; in~\SuppVisImplementation~we provide further implementation details.}


\vspace{-0.15cm}
\subsection{\change{Eclectic Network Benchmark}}
\label{sec:eclectic}

\delete{It also creates a simple metadata view showing the brief text metadata description of the input graph.}

We curated a collection of 81 networks from various sources: 24 synthetically generated graphs (derived from probabilistic edge connection models), 23 empirical networks (constructed from real-world entities and their connection measurements), 17 geometric networks (highly regular and symmetric examples from graph theory), 13 trees (topologies that lack closed loops), and 4 meshes (topologies formed from a lattice of triangles). Like Graph Atlas, all networks in this collection follow an unlabeled perspective: simple, undirected, unweighted, unlabeled, and without self-loops. \change{In~\SuppVisGrid~we provide full-page collages for each plot type supported by our pipeline, for all 81 networks.}

\change{Fig.~\ref{SelectedResults} showcases 9 representative examples, each displayed with a node-link view, and 9 of the plots generated by our visual encoding pipeline. In~\SuppVisFeature~we include high-resolution views of all plot types for each of these.}


\change{The first three networks are empirical. The first is \textit{College Football}, which consists of 115 nodes representing first-division teams (one per school) and 613 edges representing games played during the year 2000 regular season. This real-world graph captures community structure, as teams are divided into geographic conferences containing 8 to 12 teams, with more frequent games occurring within these conferences~\cite{NewmanFootball}. The second network is \textit{Bacteria Metabolism}, originally a directed graph~\cite{NetworkScienceBook}, features 1,039 nodes representing metabolites in the biochemical network of Escherichia coli, a well-known bacterial model organism. Its 4,741 edges encode metabolite-metabolite reactions. The third network is \textit{Human Metabolism}, where 2,783 nodes represent proteins within human metabolic pathways, and 6,007 edges that encode protein-protein interactions.}

\change{The next two networks are meshes, or graphs with locally similar topology (planar tilings of triangles) that can collectively construct complex topology (global 3D structure in a node-link view). The first is \textit{Stanford Bunny}, a well-known computer graphics dataset. This network has 2,503 nodes and 7,048 edges, developed in 1994 from the digital scan of a ceramic rabbit figurine~\cite{Bunny}. The second network is \textit{Fibonacci Sunflower}, a botanically inspired network we created out of 987 nodes modeled after the spatial arrangement of sunflower florets, which approximately follow the golden ratio~\cite{Vogel_1979}, from which we computed a mesh of 2,924 edges through Delaunay triangulation.}

\change{Finally, the last four networks we created with graph generator models. The first network is \textit{Stochastic Block Model}, composed of four communities of different sizes: 50, 100, 150, and 200 nodes, totaling 500 nodes. To create random intra-community density, we employ random edge wiring to connect nodes within each community 0.9 times, while nodes between communities connect only 0.1 times, resulting in a total of 42,161 edges. The second network is \textit{Erdős–Rényi}, with 1,500 nodes} \add{connected by a random 'density model'~\cite{NetworkX_ER}:} \change{node pairs are independently connected with an edge with a specified probability} \add{$p$,} \change{in our case 0.02 which results in a network of 22,512 edges. The third network, \textit{Watts-Strogatz}, also with 1,500 nodes,} \add{we generated using a 'small-world model'~\cite{NetworkX_WS}:} \change{at first, a giant ring loop connects all nodes, then each node is connected to its} \add{$k$} \change{nearest neighbors (if not already connected), and finally each edge can be moved with $p$ probability from its initial location into another random location within the network. This mix of random and ring loop edges creates a topology where otherwise distant nodes are only a few hops away from each other.} \add{We set our parameter $k$ at 32 and $p$ at 0.09, resulting in a network of 24,000 edges.} \change{The fourth network is \textit{Barabási–Albert}, also with 1,500 nodes,} \add{we generate through a preferential attachment 'scale-free model'~\cite{NetworkX_BA}:} \change{starting with a star graph, featuring a central node surrounded by} \add{$m$} \change{neighbors, the model sequentially adds new incoming nodes, each connecting with} \add{$m$} \change{edges to the existing network, with an attachment bias towards high-degree nodes. We set our parameter} \add{$m$} \change{to 16, resulting in a network of 23,744 edges.}

\vspace{-0.15cm}
\subsection{\change{Qualitative Analysis Results}}

\change{Our qualitative analysis considers the strengths, weaknesses, and redundancies between the traditional node-link views and the full suite of alternative network visualizations: Hop-Census polylines, Census-Census trajectories, and heatmap plots for each elementary constituent.} \add{We seek to understand whether the precision provided by the underlying data structures reflects in salient visual patterns. We also consider whether visually encoding different constituents yields useful information or appears redundant.} 

\add{Considering the graph topology questions of equivalence and category,} \change{in Fig.~\ref{SelectedResults} we can see that sometimes node-link views that look very similar to each other, such as \textit{Bacteria Metabolism} and \textit{Human Metabolism}, show highly visible differences in their Census plots.} \add{Across our set of 81 graphs shown in~\SuppVisGrid, we see that visibly similar families in the node-link views sometimes correspond to similarities in Census plot structure, but sometimes do not.}

\add{In some cases, all of the encodings show roughly commensurate information, albeit through different idioms. Both the Census-Census plots and the Hop-Census polyline patterns for \textit{Stochastic Block Model} show four polyline bundles, which align with the visible community structure in the node-link view and the four distinct colored regions in the heatmap plots.} \change{For the \textit{Stanford Bunny} and \textit{Fibonacci Sunflower}, the polyline overlap patterns of the Hop-Census plots and the color intensity of the Heatmap Matrix are nearly indistinguishable.} \add{These meshes have a large diameter and thus a high hop count, so there is little benefit to the polyline representation: the many parallel Hop-Census axes have such minimal display space between them that the result is visually similar to the aggregate heatmap.}

\add{In other cases including \textit{College Football}, where the hop density (graph diameter) is lower, the geometric shape of the Hop-Census polylines does visually convey more precise information than the aggregate colored boxes of Heatmap Matrix views~\cite{Lam_2007}.} \change{In particular, the inflections and peaks that} \add{result from polyline segments criss-crossing between axes} \change{provide highly salient visual patterns, just as with parallel coordinate views. For instance, the Hop-Census plots of the \textit{Bacteria Metabolism}, \textit{Human Metabolism}, \add{\textit{Erdős–Rényi}}, and \textit{Barabási–Albert} networks} \add{all have visible criss-crossing patterns. Another interesting Hop-Census pattern is the 'two-mound' structure across multiple hops in the \textit{Fibonacci Sunflower} graph: a sharp narrow peak, and a round shallower mound. Further analysis revealed sharper peak comes from core nodes at the center of the surface mesh.}

\change{All three Census trajectory plots—CN-CE, CE-CS, and CN-CS—exhibit a shared pattern that primarily differs in terms of shearing or skewing transformations. While the rough 'horseshoe' trajectory shape is consistent across Census-Census plots, the range of visible trajectory separation is greatest in CN-CS plots, slightly less in CN-CE plots, and substantially more collapsed in CE-CS plots.} The CN-CS plot shows the most salient pattern for three reasons: first, Census-Stub has the highest discerning power among these descriptors, enriching the Census-Census plot; second, Census-Node vectors always end at 0 (node degree zero), so trajectories with Census-Node on the horizontal axis always touch the vertical axis (unlike CE-CS plots, which end in the middle); and third, while Census-Node is less discerning than Census-Stub, it plays a foundational role in BFS traversal, so the combination of Census-Node and Census-Stub reveals deeper insights about topological constituents. The CE-CS plot is the least salient due to the narrow information differential between edge degree (Census-Edge) and stub degree (Census-Stub), as expected since edges are composed of only two stubs. This rationale also applies to CN-CE plots, which, though less discerning than CN-CS, still make salient the underlying Census-Census pattern. \add{In summary, CN-CS is superior to the other two trajectory plots.}

\change{Although in many cases all three Hop-Census plots are quite similar, in some cases there are visually distinguishable differences between the CN and CS plots, as visible in the generator graphs \textit{Stochastic Block Model} and \textit{Barabási-Albert}, arising from the differences in their underlying data structures. However, the CE plots do not provide much value beyond what is visible in the CS plots, since as above there is a close relationship between edges and stubs; none of the 81 benchmark graphs yielded interesting differences between those two plots \change{(\SuppVisGrid)}.} \add{We thus find that CS Hop-Census plots are superior to CE plots, but CN plots occasionally provide additional information.} 

The CN-CS plot for \textit{Bacteria Metabolism} shows about a dozen bundles of densely packed trajectories that are visually distinct from one another. \change{Further analysis indicates that this graph contains approximately six hubs (high-degree nodes), which function as topological bottlenecks. The bundles in the CN-CS plot correspond to the immediate neighbors of these hubs.} \add{Nodes that share trajectory bundles are on the same side of a particular hub bottleneck.} These hubs cannot be seen in the node-link view, where the only visible structure is a dense inner core with many peripheral tendrils. \add{This hub structure is also much less apparent in the Hop-Census polylines, which primarily highlight two peaks; only the differential trajectories in the Census-Census plots allow us to distinguish the dozen bundles. Notably, this structural feature is entirely absent in heatmap plots.} 




\begin{figure*}[!t]
\centering
\vspace{2pt}
\includegraphics[width=0.98\linewidth]{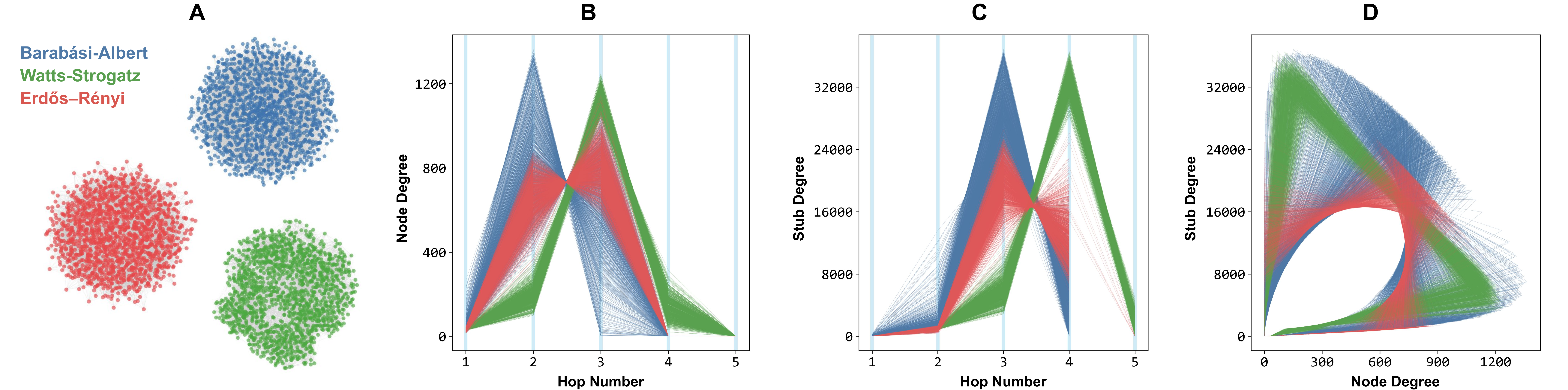}
\vspace{-3pt}
\caption{\add{Combining the three generated graphs at the bottom of Fig.~\ref{SelectedResults} as three color-coded connected components of a single disconnected graph yields distinctive superimposed Census plots. \textbf{(A)} Components in the node-link view all look like similar hairballs. \textbf{(B)} Superimposed Census-Node polylines show how Hop-Census plots afford a shared, absolute frame of reference inherited from the Census invariant descriptor. \textbf{(C)} The same absolute frame of reference holds for a Hop-Census plot of the Census-Stub instantiation. \textbf{(D)} The superimposed Census-Census plot shows the diverging structures visible in the CN-CS trajectories.}}
\vspace{-6pt}
\label{TriplePlot}
\end{figure*}


\change{In summary, these 9 networks and our rendering pipeline allowed us to inspect our entire family of Census plots, in support of a preliminary assessment of their utility; similar in spirit to the design of our quantitative combinatorial study of the full family of Census data structures.}

\delete{We find that CN-CS, the Census-Census plots where node degrees are perpendicular to stub degrees, result in the most salient patterns.}

\delete{In some cases, this Census-Census underlying pattern appears to be conserved across different networks. For example, the 'guitar pick' pattern appears for both the \textit{bacteria-metabolism} and \textit{barabasi-albert} networks. However, in \textit{bacteria-metabolism} the pattern is drawn by noisy, backdrop trajectories; while in \textit{barabasi-albert} the 'guitar pick' pattern is crisp and sharply defined. Given the presence of highly connected hubs in \textit{bacteria-metabolism}, the 'guitar pick' pattern may be a visual hallmark of networks with preferential attachment topology, which \textit{barabasi-albert} is a model of.}

\vspace{-0.15cm}
\subsection{\add{Absolute Frame of Reference Affordance}}
\label{sec:frameofreference}

\add{A strength of all Census plots is their absolute coordinate frame that allows meaningful comparisons to be made within the same plot, and across plots of different graphs. In Fig.~\ref{SelectedResults}, all plots have been individually normalized into a unit square to maximize the visual saliency of the figure. In~\SuppVisFeature~we show the native range of their axes; where within each plot type the axes do capture the same absolute counts, such as elementary constituent degree and hop number, so it can be instructive to directly compare these coordinate ranges across different graphs.}

\add{In Fig.~\ref{TriplePlot} we combine the three generated graphs showcased in Fig.~\ref{SelectedResults} into a single disconnected graph, color-coded by component source—also showcasing that BFS-Census works for disconnected graphs. While each connected component in the node-link view (Fig.~\ref{TriplePlot}-A) looks like a similar hairball, the Census polylines (Fig.~\ref{TriplePlot}-B/C) and trajectories (Fig.~\ref{TriplePlot}-D) appear distinctly different from one another, in addition to occupying mostly different regions of Census space. These superimposed plots illustrate the utility of the absolute frame of reference that Census plots exploit, where different invariant descriptors can be compared within a meaningful shared context. Notably, these three generated source graphs all have the same number of nodes (1,500), density (0.02), and a similar number of edges (about 23,400); however, they come from very different generating algorithms, and this diversity is reflected in the Census plots despite being hidden in the node-link plots.}

\section{\add{Sensitivity Analysis}}

\add{We conduct a preliminary sensitivity analysis of our approach, to consider the extent to which changes in the graph topology are proportional to the effects on our Census data structures and visual encodings. We now report on two analyses showing that large changes of topology do yield large changes of results. In~\SuppSensitivity~we include a third analysis showing that global topological sameness, despite local differences, leads to consistently similar Census plots.} 

\begin{figure*}[!htp]
\vspace{5pt}
\includegraphics[width=0.98\linewidth]{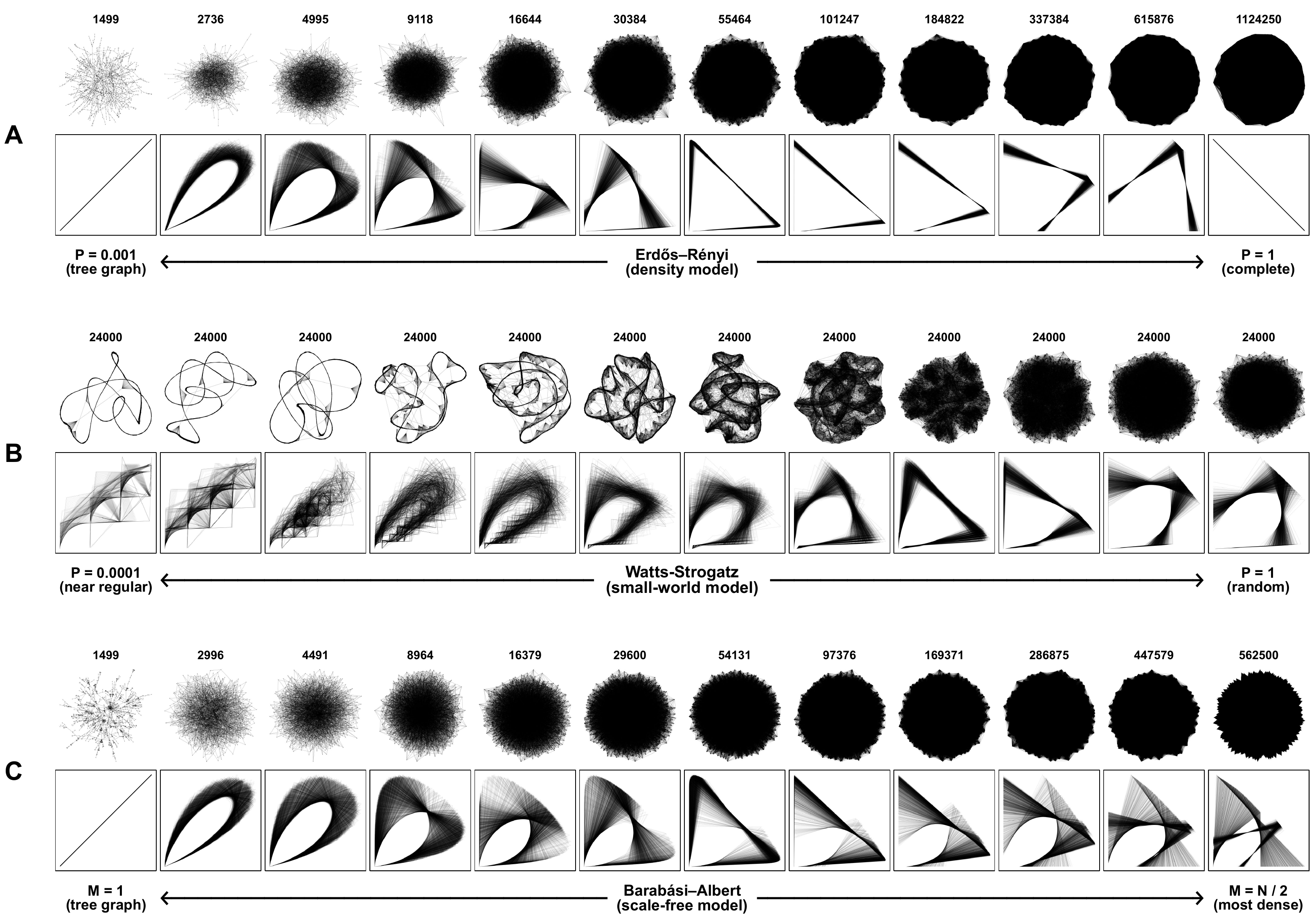}
\caption{\change{Sensitivity analysis using a continuum of 12 instances of 1500-node graphs arising from three different generator models, computed by changing the model's parameter through a geometric series. 
Each continuum has force-directed node-link view with number of edges above, and the CN-CS plot below. \textbf{(A)} Erdős–Rényi, density model.} \add{\textbf{(B)} Watts-Strogatz, small-world model. \textbf{(C)} Barabási–Albert, scale-free model.} 
\delete{Continuum of fully connected random graphs with edge densities varying from a sparse tree (left) to a fully connected graph (right), with node-link views above and CN-CS plots below. \textbf{(A)} Geometric series of fully connected graphs of 25 nodes each, with the number of edges of each reported below the node-link view. At 25 nodes none of the graphs are hairballs, so we can see density changes in the node-link view. \textbf{(B)} Same density series, but with 10 times the number of nodes: 250 nodes per graph. While the node-link views become increasingly overplotted hairballs that are difficult to distinguish from each other, the density changes are captured by the progression of CN-CS structures from a single positive-slope line through horseshoe shapes evolving from ellipsoidal to angular, with the empty eye in the middle growing until the final negative-slope line.}}
\vspace{-6pt}
\label{GeometricContinuums}
\end{figure*}

\vspace{-0.15cm}
\subsection{\add{Comparing Generator Models}}

\change{We explore} \add{three} \change{generator models through a continuum of 12 instances each, following a geometric series for model parameters. Each of these continua arises from a well-known model with predictable topological properties, so we can extend the comparison in Fig.~\ref{TriplePlot} to denser hairballs generated by the same models,} \add{for a detailed exploration of the sensitivity of Census plots.} \change{In all three cases, we see the Census-Census plots change substantially across the parameter range, whereas node-link views become indistinguishable hairballs for much of the range.} 

\change{The \textit{Erdős–Rényi} 'density model' in Fig.~\ref{GeometricContinuums}-A ranges from maximally sparse to maximally dense, with random edge wiring topology at each instance.} On the left extreme is a random tree with $n-1$ edges with a CN-CS plot that looks like a positive slope line; while on the right extreme is the complete graph with $n\choose2$ edges with a negative slope CN-CS plot. \add{This continuum is an example of major changes in graph topology yielding major changes in the Census visual encoding.} We can conceptualize these CN-CS slopes as the result of a 'phase shift' between Census-Node and Census-Stub \add{polylines}, inspired by the concept of 'phase' from signal processing—where the time domain is replaced by the hop number within a trajectory. In this analogy, the number of hops from a source node is analogous to time. When CN-CS trajectories are completely 'in-phase,' it means that for every stub collected in Census-Stub, only newly encountered nodes are collected in Census-Node. This occurs only in tree graph topologies, which lack closed loops. Visually, when Census-Node and Census-Stub are completely in-phase, their Census-Census plot forms a positive slope line. Conversely, if CN-CS are completely 'out-of-phase,' each stub collected in Census-Stub corresponds only to already visited nodes in Census-Node. This scenario arises only in complete graphs, where every node is connected to every other node, and at the first hop, all nodes are marked as visited simultaneously. Visually, when Census-Node and Census-Stub are entirely out-of-phase, their Census-Census plot forms a negative slope line.

\add{The \textit{Watts Strogatz} 'small-world model' in Fig.~\ref{GeometricContinuums}-B spans from a near-regular ring structure, to maximally random, across a continuum of rewiring probabilities without density change. This continuum illustrates how hairballs can arise without changes in density, as the right extreme is a completely randomly wired hairball that has the same number of edges as the left extreme. Again, from a sensitivity analysis point of view, we see that large changes in topology correspond to large changes in visual encoding. Furthermore, the right extreme is an \textit{Erdős–Rényi} graph, which given its 24,000 edges we can expect would be nestled between 16,644 and 30,384 edges in Fig.~\ref{GeometricContinuums}-A, and indeed, the CN-CS plots in that span look like the right extreme CN-CS plot of Fig.~\ref{GeometricContinuums}-B. In contrast, the left extreme of Fig.~\ref{GeometricContinuums}-B shows sharply angular, regular trajectories, which means the topology these are drawn from is riddled with topological bottlenecks. This is indeed the case, given that the left extreme is a near-regular \textit{Watts-Strogatz} with a handful of internal bridges that break motif regularity. We can see a continuum of how this loss in regularity evolves across Fig.~\ref{GeometricContinuums}-B, from near-regular to completely random.}

\add{The \textit{Barabási–Albert} 'scale-free model' in Fig.~\ref{GeometricContinuums}-C ranges from maximally sparse to maximally dense, like the continuum in Fig.~\ref{GeometricContinuums}-A; we again see large topological changes reflected by major visual encoding changes. However, in contrast to the \textit{Erdős–Rényi} continuum, the topology of \textit{Barabási–Albert} graphs comes from a preferential attachment model, which causes a power law distribution of node degrees within an otherwise random topology. Thus, the Fig.~\ref{GeometricContinuums}-A/C continuums start the same, with the positive slope CN-CS plots, but evolve differently towards their maximally dense extreme given the generator models are different. In Fig.~\ref{GeometricContinuums}-C a distinct, sharp bundle breaks from the main pattern in Fig.~\ref{GeometricContinuums}-A, visible from 169,371 to 562,500 edges. These trajectories come from the immediate neighbors of extremely high-degree nodes, which form the exponentially connected hubs that are a staple of the 'scale-free model'; an emergent pattern salient in the most dense CN-CS plot on the right extreme. The maximally dense preferential attachment (right extreme) is only about half of the edges of its Fig.~\ref{GeometricContinuums}-A counterpart, because in the 'scale-free model' an $m$ value above $n/2$ causes hub nodes to become maximally connected star graphs, which have fewer edges than complete graphs. Indeed, the highest possible value of $m$, at $n-1$, generates a star graph of $n$ nodes.}

\add{The previous analysis focused on model parameters that caused major topological changes. In~\SuppSensitivity, we present the opposite analysis: using the same graph generator parameters but different random conditions to create instantiations that with similar global topology (as expected from the generator model) despite having unique local differences in edge wiring. We confirm that despite these local variations, the Census plots remain consistent, with their similarity capturing the global topological patterns of graphs that belong to the same category.}

\delete{These differential patterns that Census-Census plots surface are inscrutable in the Fig.~\ref{GeometricContinuums}-A/B/C hairball node-link views. In general, while both Hop-Census and Census-Census plots experience overplotting, they are not identical and are always within the same frame of reference, because Census instantiations are invariant descriptors. In other words, Census plots translate the hairball phenomenon into a more diverse set of patterns—and crucially, in the same reliable way, which is a key analytical affordance.}

\delete{We focus on \textit{erdos-renyi}—the simplest graph generator, which places edges at random for a fixed number of nodes—to model two series of fully connected random graphs: one with 25 nodes (Fig.~\ref{Phaseshift}-A), and another with 250 nodes (Fig.~\ref{Phaseshift}-B). We visually encode a node-link view and a CN-CS plot for each graph.}

\vspace{-0.15cm}
\subsection{\add{Decreasing Graph Diameters}}

\begin{figure*}[!htp]
\centering
\vspace{2pt}
\includegraphics[width=0.98\linewidth]{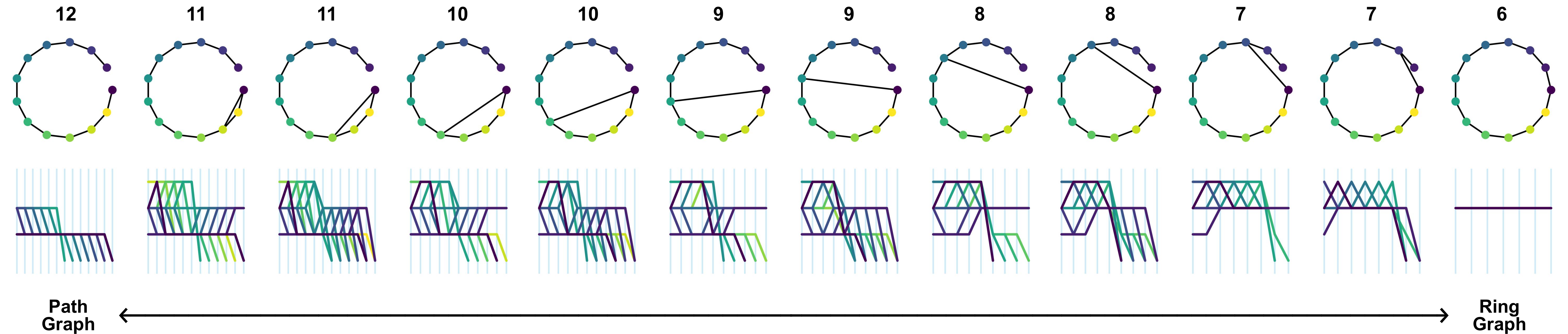}
\vspace{-3pt}
\caption{\add{Sensitivity analysis changes the diameter of a small graph of 13 nodes and edges by rewiring a single edge to cause dramatic changes to topology. At the top is the diameter, then a node-link view using a circular layout, and below is Hop-Census plots of Census-Stub polylines; both are color-coded by node index. In each Hop-Census plot, the number of horizontal axes is equal to the graph's diameter plus 1. The vertical axes are stub degrees, ranging between 0 and 3. We see that no Hop-Census plot is the same, showing sensitivity to meaningful changes in topology.}}
\vspace{-7pt}
\label{SensitivityAnalysis}
\end{figure*}

\add{In another sensitivity analysis, we change the diameter of a small graph through stepwise rewiring of an edge, as shown in Fig.~\ref{SensitivityAnalysis}. We see that meaningful changes in topology again map to conspicuous changes in the invariant descriptor and visual encoding in the horizontal extent of the Census-Stub polylines. Although the overall structure of these plots may appear similar at first glance, the pattern of the Census-Stub polyline is in fact unique to each step.}

\vspace{-0.15cm}
\section{Limitations and Future Work}

Our current implementation of BFS-Census is limited to three elementary constituents—nodes, edges, and stubs—which can be counted directly. Future work could expand these to include additional topological features, such as triangles, closed loops, and star motifs. Algorithmic techniques can involve subgraph computation inspired by GraphPrism~\cite{Kairam_2012}, which enables the computation of any whole-graph invariant descriptor by sequentially subtracting previous subgraphs, allowing for computation without the effects of cumulative collection.~\change{We designed BFS-Census with simple, unlabeled, and undirected graphs in mind,} \add{but BFS-Census works out-of-the-box for disconnected and directed graphs as well. We share an initial exploration of Census plots for the disconnected scenario in Sec.~\ref{sec:frameofreference}, and reserve the directed scenario for future work.}

While we have found substantial value in the systematic approach of our Graph Atlas Collider tool at revealing characteristics of invariant descriptors, it is important to acknowledge that some conclusions may not apply to the larger graphs found in real-world datasets, particularly since Graph Atlas caps at 10 nodes. Of course, fully systematic analyses become infeasible at larger scales because of the combinatorial explosion; there are more possible graphs with 100 nodes than atoms in the observable universe. 

\change{We have only begun to analyze the interpretation of Hop-Census} \add{polylines} \change{and Census-Census trajectories.} There is significant opportunity for future work to comprehensively explore their strengths and weaknesses, where detailed qualitative analyses are needed both within and across families of network classes, as described in Sec.~\ref{sec:eclectic}. \change{Comparing network structural hallmarks is a critical challenge in graph analysis~\cite{Tantardini_2019}, so future work calls for developing distance metrics for numeric comparison from our Census data structure, and designing specific unlabeled graph perspective tasks.} An interactive implementation that allows for bidirectional selection between Census plots and node-link views would facilitate a deeper exploration of which tasks are well supported by different visual approaches. This work could inform the design of quantitative human-subjects experiments aimed at better assessing the utility of Census plots.

\delete{We show the changes in CN-CS across the density extremes of a random graph generator model (Fig.~\ref{Phaseshift}); future work can explore such Census-Census continuum plots for other generator models, such as preferential attachment.}

\delete{Results from our Graph Atlas Collider testing tool (Fig.~\ref{collisions_TGC} and Fig.~\ref{TGC_Results}) reveal regions of Graph Atlas that Census-Stubs cannot discern. These 'dark regions' are collision sets. Future work can probe these sets to further understand the characteristics of graph topology that affect invariant descriptor discerning power. Our Collider tool is based on the Graph Atlas benchmark, which is comprehensive but limited to only tiny graphs.}

\vspace{-0.15cm}
\section{Conclusion}

\change{
In this paper we described a general approach for invariant graph desciptors (BFS-Census) and an analysis of their discerning power. We also proposed visual representations of graphs based on the invariant descriptors and demonstrated the results on synthetic and real-world graphs.}

\change{The BFS-Census algorithm computes the Census family of invariant descriptors for nodes, edges, and stubs—three elementary constituents of graph topology. Our experiments show that Census-Stub has orders of magnitude more discerning power, yet minimal computing power and storage space cost. }
\change{
The visual encoding of Census trajectories, such as the Hop-Census and Census-Census, allow us to gain insight into topological properties through the comparison of the visual encoding of the discerning invariant descriptors.}

\vspace{-0.15cm}
\section*{Supplemental Materials Index}

Supplemental materials are available at \textit{\href{https://osf.io/nmzra/}{https://osf.io/nmzra/}}.

\change{We provide a supplemental PDF with five sections: the algorithmic details of the previous network portrait work including illustrated explainers of the BFS-BMatrix}
\add{and GraphPrism approaches}
\change{(\SuppOne); implementation details of our BFS-Census algorithm (\SuppTwo); further discussion of our quantitative Graph Atlas Collider tool}
\add{including an illustrated example of a 20-node collision noted in previous work}
\change{(\SuppThree); further details of our qualitative analysis through a visual encoding pipeline (\SuppFour)—including many additional high-resolution images: full-page views for each of the networks featured in this paper (\SuppVisFeature), and full-page collages for each plot type across all 81 networks (\SuppVisGrid);}
\add{and an additional sensitivity analysis results for 81 generated graphs that share topological similarity despite local differences in edge wiring, yet Census plots consistently capture their global similarity (\SuppFive).}

We also provide a supplemental ZIP file with open-source code and analysis scripts for all algorithms and software discussed in this paper, input data, input data provenance details, all Collider output data, and all result images from the visual encoding pipeline.

\vspace{-0.2cm}
\section*{Acknowledgements}

The development of this paper took place between September 2022 and October 2024 at the UBC Point Grey campus and in the City of Vancouver, which sit on the traditional, ancestral, unceded territory of the Musqueam, Squamish, and Tsleil-Waututh First Nations. This work was supported in part by NSERC DG RGPIN-2014-06309. We thank Daniel Farmer and Robert Xiao for their software engineering advice; Steve Kasica, Francis Nguyen, Ryan Smith, and Mara Solen for their manuscript reviews; Laks Lakshmanan and Daniel Weiskopf for their advice on network theory; \add{and TVCG reviewers for their thoughtful, constructive feedback that substantially strengthened this paper}.


\bibliographystyle{IEEEtran}
\bibliography{IEEEabrv, references}

 \vskip -2\baselineskip plus -1fil

\begin{IEEEbiography}[{\includegraphics[width=1in,height=1.25in,clip,keepaspectratio]{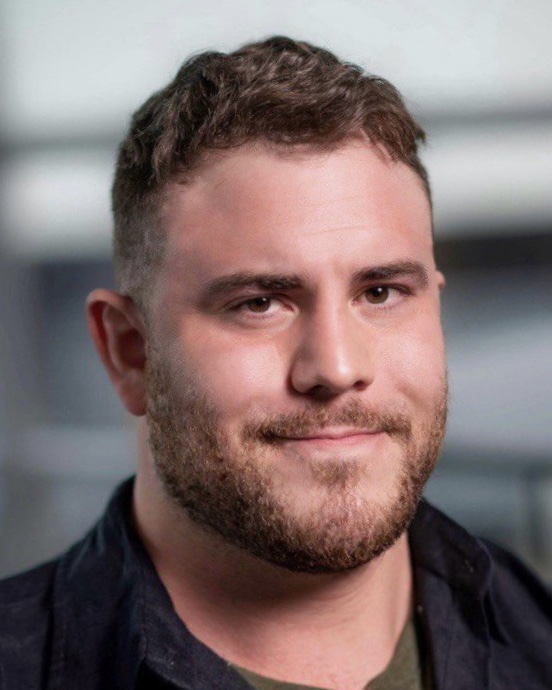}}]{Matt I.B. Oddo} Matías Ignacio Bofarull Oddó (Matt Oddo) is a PhD student in information visualization at the Department of Computer Science of the University of British Columbia since 2022. He moved to Canada from Chile in 2011. He received the BSc degree in biology from UBC-Okanagan in 2016 and the MSc degree in earth science from UBC in 2019. He is a data scientist while not at UBC, at Lucent Biosciences.
\end{IEEEbiography}

 \vskip -2\baselineskip plus -1fil

\begin{IEEEbiography}[{\includegraphics[width=1in,height=1.25in,clip,keepaspectratio]{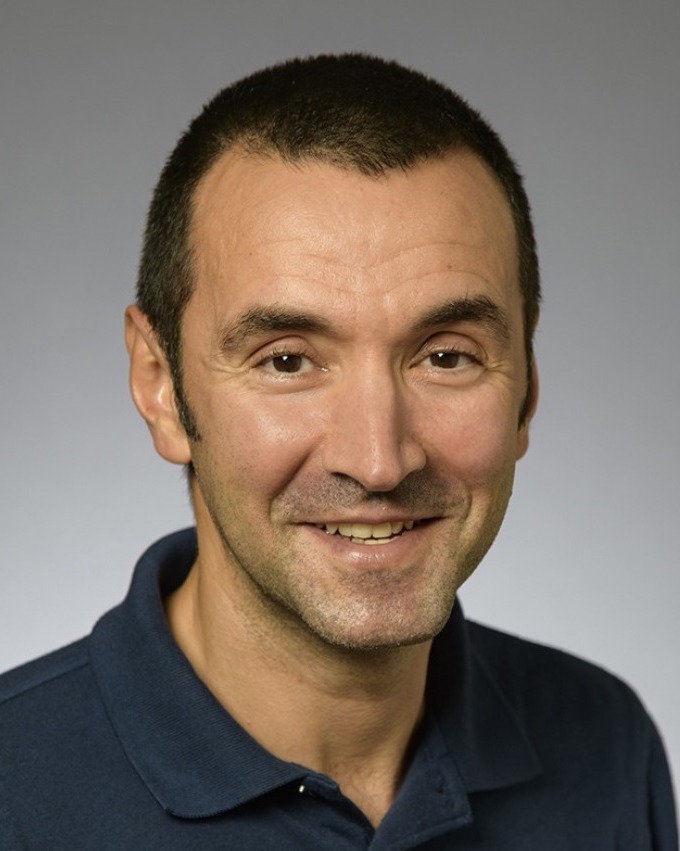}}]{Stephen Kobourov} is a Professor of Computer Science at the Technical University of Munich. He received  BS degrees in Mathematics and Computer Science from Dartmouth College and MS and PhD
degrees from Johns Hopkins University. His research interests include information visualization, graph theory, and geometric algorithms. 
\end{IEEEbiography}

 \vskip -2\baselineskip plus -1fil
 
\begin{IEEEbiography}[{\includegraphics[width=1in,height=1.25in,clip,keepaspectratio]{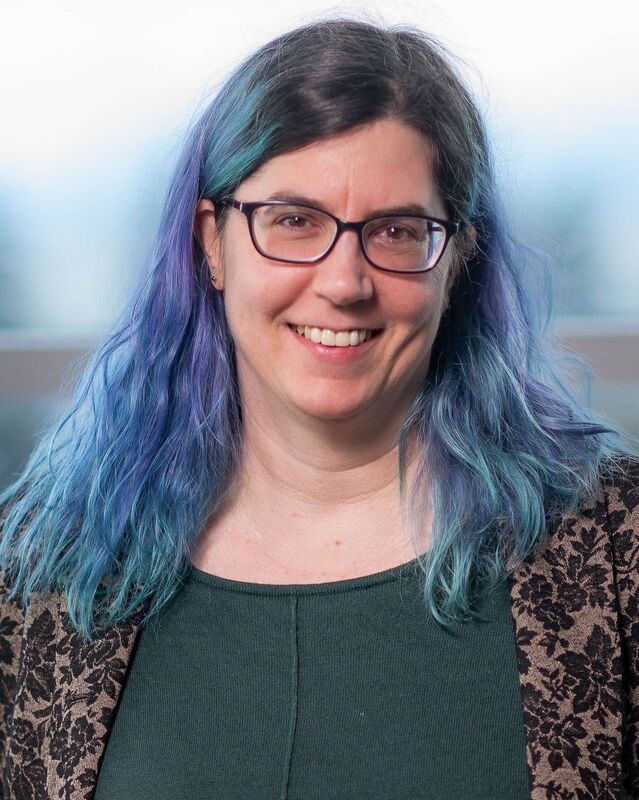}}]{Tamara Munzner}
(IEEE Fellow) received the PhD degree from Stanford. She is currently a professor with the University of British Columbia. She has worked on visualization projects in a broad range of application domains from genomics to journalism. Her book Visualization Analysis and Design is heavily used worldwide, and she was the recipient of the IEEE VGTC Visualization Technical Achievement Award.
\end{IEEEbiography}

\end{document}